\documentclass[twocolumn,english,aps,pra,10pt,superscriptaddress,floatfix]{revtex4-2}
\usepackage{times}
\usepackage{graphicx}
\usepackage{amssymb}
\usepackage{bm}% bold math
\usepackage{amssymb,amsfonts,amsmath,amsbsy,bm,t1enc,latexsym}
\usepackage{float}
\usepackage[colorlinks=true,citecolor=blue,linkcolor=magenta]{hyperref}
\usepackage[markup=blue, authormarkupposition=left]{changes}
\usepackage[english]{babel}
\usepackage{url}
\usepackage{siunitx}
\DeclareSIUnit \dbc {dBc}
\usepackage{soul}
\usepackage{changes}
\usepackage{cancel}
\usepackage{times}
\newcommand{\Fref}[1]{Figure~\ref{#1}}

\newcommand{\SiN}[0]{$\mathrm{Si}_3\mathrm{N}_4$}
\newcommand{\LN}[0]{$\mathrm{LiNbO}_3$} 
\newcommand{\LT}[0]{$\mathrm{LiTaO}_3$}

\newcommand*{\bigchi}{\mbox{\Large$\chi$}}% big chi

\begin{document}
%\title{Pyroelectrically modified thermal response and intrinsic cavity noise in ferroelectric photonics}
%\title{Pyro-Electro-Optic effects and Flicker noise in ferroelectric integrated photonics.}
%\title{Johnson Noise Limits Pockels Integrated Photonics}
\title{Fundamental charge noise in electro-optic photonic integrated circuits}
%\title{Fundamental charge noise in ferroelectric photonic integrated circuits}
%\title{Johnson Noise Limits Pockels Integrated Photonics}

\author{Junyin Zhang}
\affiliation{Institute of Physics, Swiss Federal Institute of Technology Lausanne (EPFL), CH-1015 Lausanne, Switzerland}
\affiliation{Center of Quantum Science and Engineering, EPFL, CH-1015 Lausanne, Switzerland}

\author{Zihan Li}
\affiliation{Institute of Physics, Swiss Federal Institute of Technology Lausanne (EPFL), CH-1015 Lausanne, Switzerland}
\affiliation{Center of Quantum Science and Engineering, EPFL, CH-1015 Lausanne, Switzerland}

\author{Johann Riemensberger}
\email[]{johann.riemensberger@epfl.ch}
\affiliation{Institute of Physics, Swiss Federal Institute of Technology Lausanne (EPFL), CH-1015 Lausanne, Switzerland}
\affiliation{Center of Quantum Science and Engineering, EPFL, CH-1015 Lausanne, Switzerland}

\author{Grigory Lihachev}
\affiliation{Institute of Physics, Swiss Federal Institute of Technology Lausanne (EPFL), CH-1015 Lausanne, Switzerland}
\affiliation{Center of Quantum Science and Engineering, EPFL, CH-1015 Lausanne, Switzerland}

\author{Guanhao Huang}
\affiliation{Institute of Physics, Swiss Federal Institute of Technology Lausanne (EPFL), CH-1015 Lausanne, Switzerland}
\affiliation{Center of Quantum Science and Engineering, EPFL, CH-1015 Lausanne, Switzerland}

\author{Tobias J. Kippenberg}
\email[]{tobias.kippenberg@epfl.ch}
\affiliation{Institute of Physics, Swiss Federal Institute of Technology Lausanne (EPFL), CH-1015 Lausanne, Switzerland}
\affiliation{Center of Quantum Science and Engineering, EPFL, CH-1015 Lausanne, Switzerland}

\maketitle

%\section*{Abstract}
\textbf{
	Understanding thermodynamical measurement noise is of central importance for electrical and optical precision measurements from mass-fabricated semiconductor sensors, where the Brownian motion of charge carriers poses limits\cite{johnsonThermalAgitationElectricity1928,nyquistThermalAgitationElectric1928}, to optical reference cavities for atomic clocks or gravitational wave detection, which are limited by thermorefractive and thermoelastic noise due to the transduction of temperature fluctuations to the refractive index and length fluctuations\cite{levinInternalThermalNoise1998,LEVIN20081941,liuThermoelasticNoiseHomogeneous2000}.
    Here, we discover that unexpectedly charge carrier density fluctuations give rise to a novel noise process in recently emerged electro-optic photonic integrated circuits.
    We show that Lithium Niobate and Lithium Tantalate photonic integrated microresonators exhibit an unexpected Flicker type (i.e. $1/f^{1.2}$) scaling in their noise properties, significantly deviating from the well-established thermorefractive noise theory\cite{levinInternalThermalNoise1998,huangThermorefractiveNoiseSiliconnitride2019,panuskiFundamentalThermalNoise2020}. 
    We show that this noise is consistent with thermodynamical charge noise\cite{brunsThermalChargeCarrier2020,siegelRevisitingThermalCharge2023}, which leads to electrical field fluctuations that are transduced via the strong Pockels effects of electro-optic materials. 
    Our results establish electrical Johnson-Nyquist noise as the fundamental limitation for Pockels integrated photonics, crucial for determining performance limits for both classical and quantum devices, ranging from ultra-fast tunable and low-noise lasers, Pockels soliton microcombs, to quantum transduction, squeezed light or entangled photon-pair generation\cite{wangIntegratedLithiumNiobate2018,snigirevUltrafastTunableLasers2023,liIntegratedPockelsLaser2022,heSelfstartingBichromaticLiNbO2019,nehraFewcycleVacuumSqueezing2022,stokowskiIntegratedQuantumOptical2023,sahuEntanglingMicrowavesLight2023,zhaoHighQualityEntangled2020,xuBidirectionalInterconversionMicrowave2021}. Equally, this observation offers optical methods to probe mesoscopic charge fluctuations with exceptional precision.
}

Thermal noise plays a crucial role in various fields, ranging from the electronics \cite{johnsonThermalAgitationElectricity1928,nyquistThermalAgitationElectric1928,vossFlickerNoiseEquilibrium1976} and semiconductor industry \cite{suryaTheoryExperimentNoise1986,hungUnifiedModelFlicker1990, jimminchangFlickerNoiseCMOS1994,jayaramanNoiseTechniqueExtract1989}, to photonics and metrology, i.e. reference cavities for atomic clocks and gravitational wave detection \cite{ludlow2007compact,huangThermorefractiveNoiseSiliconnitride2019,gorodetskyFundamentalThermalFluctuations2004,panuskiFundamentalThermalNoise2020,levinInternalThermalNoise1998,LEVIN20081941}.
For optical interferometer or cavity-based measurement as used in gravitational wave detection, two key noise contributions have been established, i.e. the thermo-refractive noise and the thermo-elastic noise, which transduce stochastic temperature fluctuation via the thermo-optic or thermo-elastic coefficient to frequency fluctuations, thereby limiting precision measurements\cite{levinInternalThermalNoise1998,liuThermoelasticNoiseHomogeneous2000,matskoWhisperinggallerymodeResonatorsFrequency2007}. 
Indeed, in integrated photonics resonators, such thermo-refractive noise has been identified to limit the precision of optomechanical displacement measurements \cite{anetsbergerMeasuringNanomechanicalMotion2010} and also been observed to limit the linewidth of self-injection locked lasers based on low-loss photonic integrated circuits \cite{jinHertzlinewidthSemiconductorLasers2021,kondratievSelfinjectionLockingLaser2017}.
%Note: Here I want to focus on one special case of Flicker noise: the charge-induced Flicker noise. So Clark's work is not very suited for this kind of Flicker noise. For charge-induced Flicker noise, the contribution is from various papers instead of several big names.
In the electrical domain, it is well-established that the Brownian motion of charge carriers induces a voltage noise at the terminals of a resistor, i.e., Johnson-Nyquist noise\cite{johnsonThermalAgitationElectricity1928,nyquistThermalAgitationElectric1928}
Over the past few decades, there has been significant focus on the investigation of charge-carrier noise beyond Johnson's model, particularly the \(1/f\) noise, often referred to as Flicker noise. 
This noise commonly features a frequency scaling represented by $S_{\nu}\propto f^{-\eta}$, with $\eta$ ranging from 0.5 to 1.5 \cite{hungUnifiedModelFlicker1990, jimminchangFlickerNoiseCMOS1994,jayaramanNoiseTechniqueExtract1989}.
Despite Flicker-type noise having been studied extensively in condensed matter physics\cite{hungUnifiedModelFlicker1990, jimminchangFlickerNoiseCMOS1994,jayaramanNoiseTechniqueExtract1989,heritierSpatialCorrelationFluctuating2021}, and having found to limit the coherence of superconducting quantum circuits\cite{kuhlmannChargeNoiseSpin2013}, NV-centers\cite{kimDecoherenceNearSurfaceNitrogenVacancy2015}, Rydberg atoms \cite{carterCoherentManipulationCold2013}, such $1/f$ type charge noise has no counterpart in optics, i.e., has never been observed in photonics or optical measurements to the best of our knowledge.
Recently, thermal-charge-carrier-refractive noise (TCCR) noise \cite{brunsThermalChargeCarrier2020,siegelRevisitingThermalCharge2023} was theoretically investigated in  mirrors for gravitational wave detectors, yet the magnitude of the refractive index change via the free-carrier dispersion effect in silicon was found too small for experimental observation.

%a novel noise process has been theoretically proposed to arise from the transduction of thermodynamical charge fluctuations to refractive index changes , termed as 
% However,  and thus will not be a limiting factor for future gravitational wave observatories\cite{brunsThermalChargeCarrier2020,siegelRevisitingThermalCharge2023}, given the weak electro-optical effect in silicon.
%This perspective was recently studied in the context of gravitational wave detection\cite{siegelRevisitingThermalCharge2023,brunsThermalChargeCarrier2020}, where the impact of Johnson Noise induced refractive noise, termed as thermal-charge-carrier-refractive noise (TCCR) in high-purity silicon optics are theoretically investigated to order smaller than other noises and thus can be omitted. 

\begin{figure*}[htbp]
 	\centering
 	\includegraphics[width=1\linewidth]{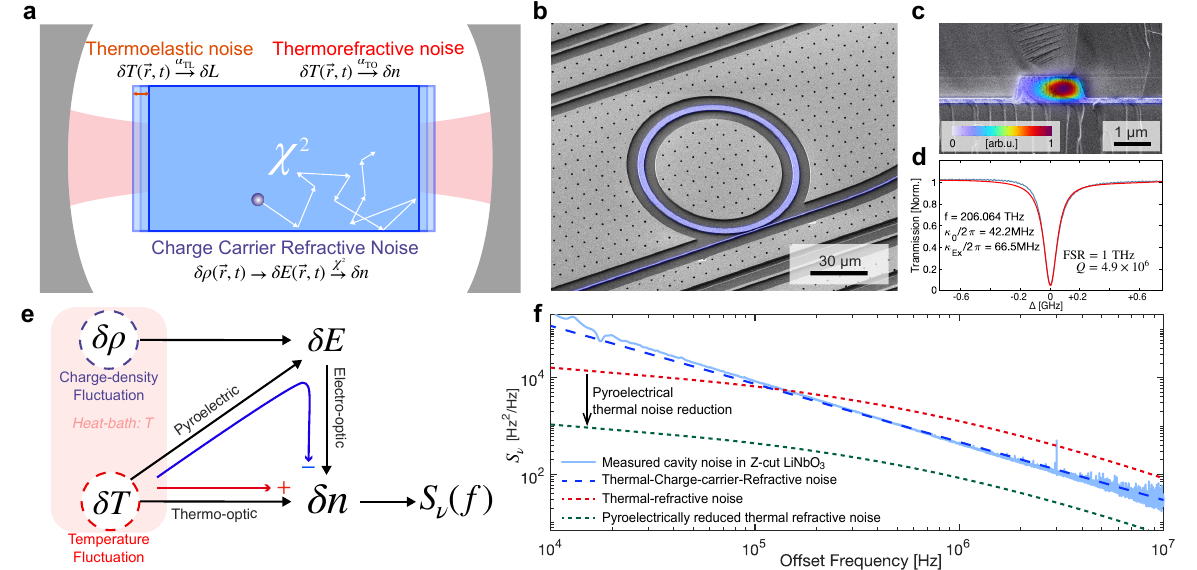}
     \caption{\textbf{Noise and thermal response in electro-optic microresonators.}
         (a) Schematic representation of noise processes in electro-optic materials. The inherent thermodynamical temperature fluctuations $\delta T$ give rise to two distinct types of noise: thermoelastic noise and thermorefractive noise (TRN). Brownian motion of charge carriers induces charge-density fluctuations $\delta \rho$, which generate electric field fluctuations that drive thermal-charge-carrier-refractive noise (TCCR) through the Pockels effect.
         (b) Scanning Electron Microscopy (SEM) image of Z-cut Lithium Niobate (\LN) devices used in the experiment with a free-spectral range of 1~THz.
         (c) SEM of the waveguide ring cross-section of the resonator. The fundamental optical mode profile is indicated in a color overlay.
         (d) Normalized transmission for 1~THz Z-cut \LN microresonator. Blue trace: measurement; Red trace: Lorentzian fit.
         (e) Schematic of the refractive noise and thermal response in electro-optic photonics.
         The temperature change will induce electric fields due to the pyroelectric effect, which then changes the refractive index via the Pockels effect. 
         The combined effect, referred to as PyroEO (Pyroelectric-Electro-Optic), contributes to a net negative thermo-optic response that reduces the overall response.
         (f) Schematic of intrinsic cavity noise in electro-optic microresonators. 
     }
 \label{fig1}
 \end{figure*}

Here we report the discovery that charge noise plays a key role in electro-optic integrated photonics. 
Electro-optic photonic integrated circuits\cite{boesLithiumNiobatePhotonics2023}, based on ferroelectric materials such as Lithium Niobate\cite{zhuIntegratedPhotonicsThinfilm2021,boesLithiumNiobatePhotonics2023}, Lithium Tantalate\cite{wang2023lithium}, or Barium Titanate\cite{abelLargePockelsEffect2019}, have widespread applications ranging from quantum optics to broadband optical communication such as ultrafast tunable lasers, optical comb generation, squeezed light generation and quantum transduction\cite{wangIntegratedLithiumNiobate2018,snigirevUltrafastTunableLasers2023,liIntegratedPockelsLaser2022,heSelfstartingBichromaticLiNbO2019,nehraFewcycleVacuumSqueezing2022,stokowskiIntegratedQuantumOptical2023,sahuEntanglingMicrowavesLight2023,zhaoHighQualityEntangled2020,xuBidirectionalInterconversionMicrowave2021} due to their strong Pockels effect and tight optical confinement that have led to a two order-of-magnitude increase in the electro-optical conversion efficiency compared with bulk devices\cite{wangIntegratedLithiumNiobate2018} while retaining ultra-low optical loss \cite{zhang2017monolithic,liTightlyConfiningLithium2022}. 
Thus far, their intrinsic thermodynamical noises, which establish the ultimate performance threshold for those applications, have not been examined.
In the following, we show that ferroelectric microresonators exhibit an unusual noise, which scales with $S_{\nu}\propto f^{-\eta}$, with $\eta=1.2$ across the whole measurement frequency range from 10~kHz to 10~MHz and which significantly deviates from the conventional prediction based on thermo-refractive noise theory. 
This noise is shown to agree with charge density fluctuations caused by the finite conductivity and transduced via the Pockels effect to refractive index fluctuations (\Fref{fig1}(a)).
Moreover, we show that the thermo-refractive noise, which is the conventionally acknowledged cavity frequency noise limit in dielectrics and semiconductors\cite{LEVIN20081941,liuThermoelasticNoiseHomogeneous2000,gorodetskyFundamentalThermalFluctuations2004,matskoWhisperinggallerymodeResonatorsFrequency2007,huangThermorefractiveNoiseSiliconnitride2019,panuskiFundamentalThermalNoise2020}, is coherently canceled due to the effective thermo-optic response caused by the combination of pyroelectric and electro-optic nonlinearities (PyroEO), making charge noise the predominant noise (\Fref{fig1}(f)) .

% %%%%%%%%%%%%%%%%%%%%%%%%%%%%%%%%%%%%%%%%%%%%%%%%%%%%%%%%%%%%%%%%%%%%%%%%%%%%%%%%%%%%%%%%%%%%
% %%%%%%%%%%%%%%%%%%%%%%%%%%%%%%% I N T R O D U C T I O N  %%%%%%%%%%%%%%%%%%%%%%%%%%%%%%%%%%%
% %%%%%%%%%%%%%%%%%%%%%%%%%%%%%%%%%%%%%%%%%%%%%%%%%%%%%%%%%%%%%%%%%%%%%%%%%%%%%%%%%%%%%%%%%%%%

% \section{Introduction}

%%%%%%%%%%%%%%%%%%%%%%%%%%%%%%%%%%%%%%%%%%%%%%%%%%%%%%%%%%%%%%%%%%%%%%%%%%%%%%%%%%%%%%%%%%%%
%%%%%%%%%%%%%%%%%%%%%%%%%%%%%%%%%%%% PyroEO       %%%%%%%%%%%%%%%%%%%%%%%%%%%%%%%%%%%%%%%%%
%%%%%%%%%%%%%%%%%%%%%%%%%%%%%%%%%%%%%%%%%%%%%%%%%%%%%%%%%%%%%%%%%%%%%%%%%%%%%%%%%%%%%%%%%%%%

\section*{Pyroelectric Coherent Thermorefractive Noise Cancellation}
\begin{figure*}[htbp]
    {
        \centering
        \includegraphics[width=1\linewidth]{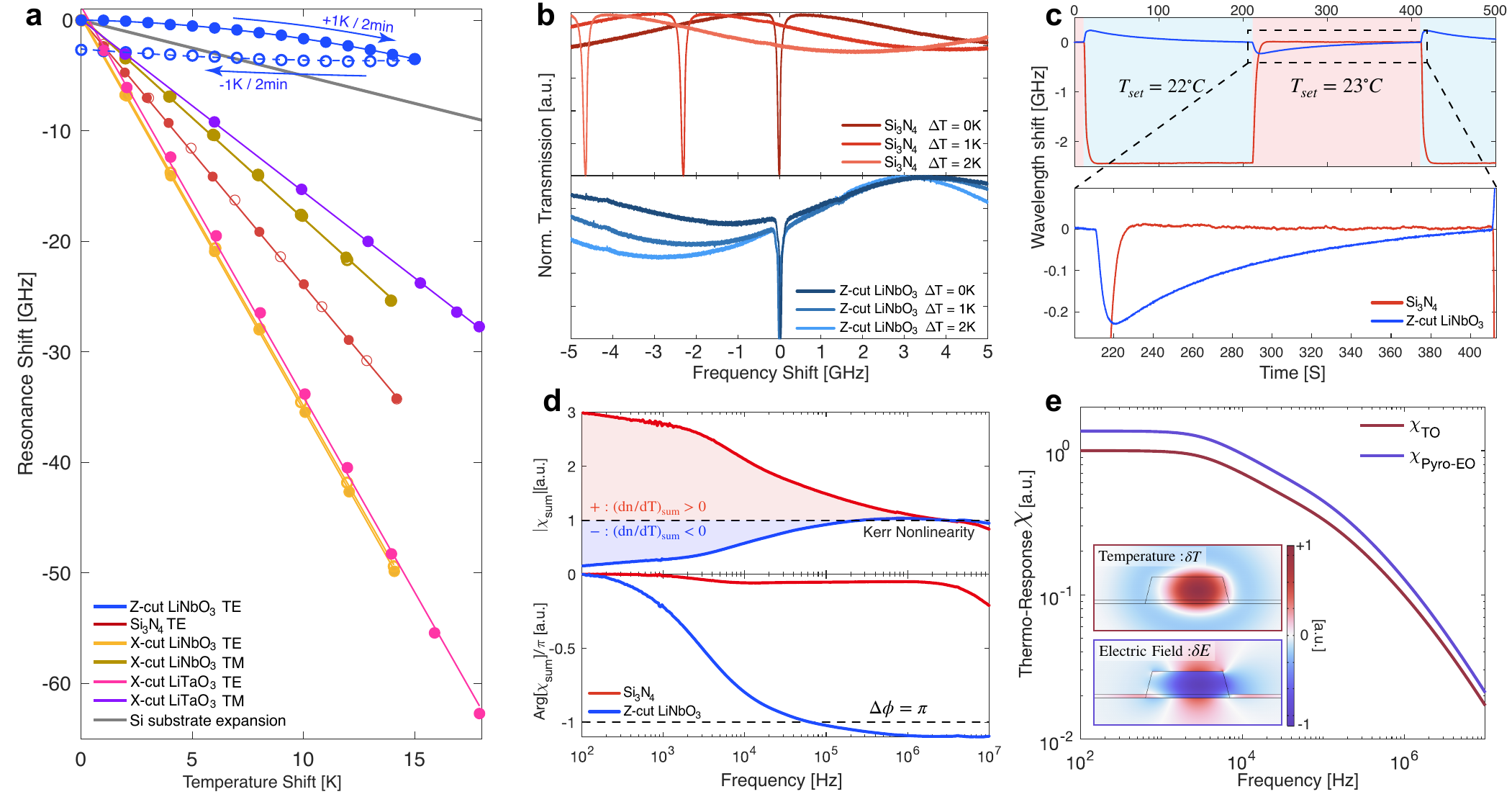}
        %\captionsetup{justification=raggedright,singlelinecheck=false}
        \caption{\textbf{Pyroelectricity modified thermo-optic response.}
        (a) Thermo-optic frequency shift of ferroelectric (\LN,\LT) and non-ferroelectric (\SiN) microresonators with 0.5~K/min heating/cooling rate. 
        Solid dots: measurement points during temperature ramp-up;
        Hollow dots: measurement points during temperature ramp down;
        Solid lines: linear fit for temperature ramp up (quadratic fit for Z-cut \LN);
        Dashed lines: linear fit for temperature ramp down (quadratic fit for Z-cut \LN);
        The hysteresis in Z-cut \LN~originates from the slow decaying nature of PyroEO in the experiment time scale.
        (b) Temperature-induced resonance shift for \SiN- and Z-cut \LN-based microresonator. 
        The normalized transmission curve is measured by the DC-shift measurement setup with 100~s delay.
        (c) Temperature-induced resonance shift ring down. The temperature is ramped down and up between 22~$^{\circ}$C and 23~$^{\circ}$C with period 400~s.
        % , during which the resonance shift is monitored by DC-shift measurement setup. 
        % The ring-down behavior measured in Z-cut Lithium Niobate indicates the existence of a pyroelectricity-induced EO effect, which can be modulated fast but decays slowly with a time constant of 100~s.
        (d) Amplitude and phase of nonlinear response normalized to Kerr nonlinearity. The phase inversion indicates a negative net thermo-optic effect due to PyroEO.
        % The abnormal response curve measured in Z-cut \LN~is due to the PyroEO which surpasses the thermal-optics nonlinearity and shifts the resonance in an opposite direction compared with Kerr in low-frequency. Such inversion induces a $\pi$-phase shift in the total response.
        (e) Finite-element simulation of thermo-optic response and PyroEO effects in Z-cut \LN-based microresonator.
        % demonstrates the frequency dependency similarity in the modified  of  Lithium Niobate, exhibiting comparable trends between the thermo-optic and  contributions.
} 
\label{fig2}
    }
\end{figure*}
We studied the cavity noise in ferroelectric optical microresonators fabricated in X-cut and Z-cut \LN~and X-cut \LT~ optical microresonators using a diamond-like-carbon (DLC) based hardmask etching process~\cite{liTightlyConfiningLithium2022,wang2023lithium} with free spectral range (FSR) between 80~GHz and 1~THz (see \Fref{fig1} (b,c)) and intrinsic optical linewidth as low as 42.2~MHz shown in \Fref{fig1}(d). 
The resonators with high $Q/V_{\mathrm{eff}}$ (quality factor/effective volume) ratios enable accurate detection of intrinsic refractive index noise.
Ferroelectric materials possess inherent dipole moments that can be modulated by external electric fields $\delta E$, changing the refractive index $n$ through the Pockels effect: $\delta n = -n^3r/2\  \delta E$, where $r$ is the Pockels coefficient in the specific polarization.
Temperature can also affect these materials' dipole moments, creating internal electric fields known as the pyroelectric effect \cite{pengOriginPyroelectricityLiNbO2011}.
This field can modify the refractive index through the electro-optic response \cite{bulmerPyroelectricEffectsLiNbO1986}, introducing a thermo-optic nonlinearity that can compensate the conventional thermo-optic response (cf.~\Fref{fig1}(e)) and thermorefractive noise.
% After this pyroelectrical thermorefractive noise reduction, the thermal-charge-carrier-noise becomes the dominating noise source (cf.~\Fref{fig1}(e,f)).

The total temperature induced refractive index change $\delta n_\mathrm{sum}$ can be described as follows:
\begin{equation}\tag{1}
    \begin{aligned}
\delta n_\mathrm{sum} &= \delta n_\mathrm{TO} + \delta n_\mathrm{PyroEO}
\approx \alpha_\mathrm{TO} \delta T + \frac{n^3rp}{2\epsilon_\mathrm{r} \epsilon_0}\delta T,
    \end{aligned}
\label{PyroEO-TO}
\end{equation}
where $\delta n_\mathrm{TO}$ refers to the conventional thermo-optic response, whereas $\delta n_{\mathrm{PyroEO}}$ denotes the effective thermo-optic response of the combination of pyroelectricity and the electro-optic effect. 
$\delta T$ is the temperature fluctuations, $\alpha_\mathrm{TO}$ is the thermorefractive coefficient. 
The PyroEO response is hence determined by the pyroelectric coefficient $p$, the Pockels coefficient $r$, the refractive index $n$, and the relative permittivity $\epsilon_\mathrm{r}$. 
% The detailed calculation process requires the finite element method to account for the distortion of the pyroelectric electric field caused by the waveguide geometry.
Coherent cancellation of the thermo-optic response is possible if the strength of the PyroEO and the thermo-optic effects match closely, which is the case for the TE-mode in z-Cut \LN waveguides (cf. supplementary materials). 
We compare the properties of ferroelectric optical microresoantors to \SiN~microresonators with FSRs from 10~GHz to 200~GHz fabricated using the photonic Damascene process \cite{liuHighyieldWaferscaleFabrication2021} and featuring comparable optical linewidth and optical mode volume. 

First, we studied the resonance shift directly caused by environment temperature change.
\Fref{fig2}(a) presents a comparison of resonance shifts for different microresonators across a 15~K temperature range above the 294~K lab temperature with a ramp rate of $\pm$0.5~K/min. 
Z-cut \LN (dark blue) is distinguished by its athermal behavior compensating the thermal expansion of the silicon substrate (solid black line) and its hysteresis, which indicates the presence of a slow-decaying pyroelectric process (for the detailed discussion see supplementary material).
\Fref{fig2}(b) depicts laser frequency scans across the resonances of a Z-cut \LN~resonantor and a \SiN~resonator for modest temperature changes of 1-2~K. 
While the \SiN~resonantor drifts by 2.3~GHz/K, the drift of the Z-cut \LN~resonator is less than 100~MHz/K featuring a more than one order of magnitude reduction in thermal drift due to the PyroEO effect. 
% In all cases, unless explicitly noted otherwise, we couple to the fundamental TE$_{00}$ mode of the optical microring.
\Fref{fig2}(c) plots the resonance shift in response to temperature switching between 22$^{\circ}$C and 23$^{\circ}$C with a period of 400~s.
The immediate resonance shift that occurs at the commencement of the temperature change, is followed by a slower decay with a time constant of around $10^2$~s (cf. supplementary material). 
% More measurement results for long-time shifting are available in the supplementary material.
The modulation of pyroelectric charges is a result of an instantaneous polarization change following atomic scale relaxation \cite{pengOriginPyroelectricityLiNbO2011}, while the decay of these charges results from surface current and charge neutralization process that involves ions from the surroundings and is typically much slower \cite{bulmerPyroelectricEffectsLiNbO1986}.
Our observations for Z-cut \LN~ substantiate that though the original thermo-optic coefficient for o-light in \LN~is positive (shown in the measurement of TM-mode in X-cut \LN), the PyroEO effect contributes to a net negative thermo-optic coefficient in Z-cut \LN (cf. Equation \ref{PyroEO-TO}).
We attribute the absence of hysteresis during our measurement in X-cut \LN~to the much faster pyroelectric decay times due to leakage current at the ion beam etched surface.

Secondly, we conduct a coherently-driven optical intensity modulation response measurement, where the temperature changes are driven by the optical absorption in the resonator directly via the photothermal effect \cite{gaoProbingMaterialAbsorption2022}.
% This experiment allows us to explore the thermo-optic nonlinearity where the heat source is the photothermal effect.
\Fref{fig2}(d) presents the nonlinear response\cite{gaoProbingMaterialAbsorption2022} of two microresonators: the non-ferroelectric \SiN microresonator and the ferroelectric Z-cut \LN microresonator. In the latter, the PyroEO effect induces a negative thermo-optic coefficient, which over-compensates the intrinsic positive thermo-optic coefficient, leading to a $\pi$-phase shift in the response phase.
% Importantly, it shows that the modulation of the PyroEO is a rapid process that can closely follow the thermal response and is not constrained by the slow decay associated with the leakage current.
More data measured at different temperatures are available in the supplementary material.
We simulated the thermal response curve for the thermo-optic effect and the PyroEO effect and depict the result in \Fref{fig2}(f). 
Both thermo-optic response and PyroEO response show similar frequency scaling from 1~kHz to 10~MHz determined by heat diffusion.
The panel insets illustrate the normalized profile of the photothermal induced temperature field $\delta T(\vec{r})$ at a frequency of 1~MHz and the corresponding electric field in Z-direction $\delta E(\vec{r},t)$ induced by the pyroelectric charge.

%%%%%%%%%%%%%%%%%%%%%%%%%%%%%%%%%%%%%%%%%%%%%%%%%%%%%%%%%%%%%%%%%%%%%%%%%%%%%%%%%%%%%%%%%%%%
%%%%%%%%%%%%%%%%%%%%%%%%%%%%%%%%%%%% Cavity Frequency Noise  %%%%%%%%%%%%%%%%%%%%%%%%%%%%%%%
%%%%%%%%%%%%%%%%%%%%%%%%%%%%%%%%%%%%%%%%%%%%%%%%%%%%%%%%%%%%%%%%%%%%%%%%%%%%%%%%%%%%%%%%%%%%

\section*{Fundamental Charge Refractive Noise via Pockels effect}
\begin{figure*}[htbp]
    {
        \centering
        \includegraphics[width=1\linewidth]{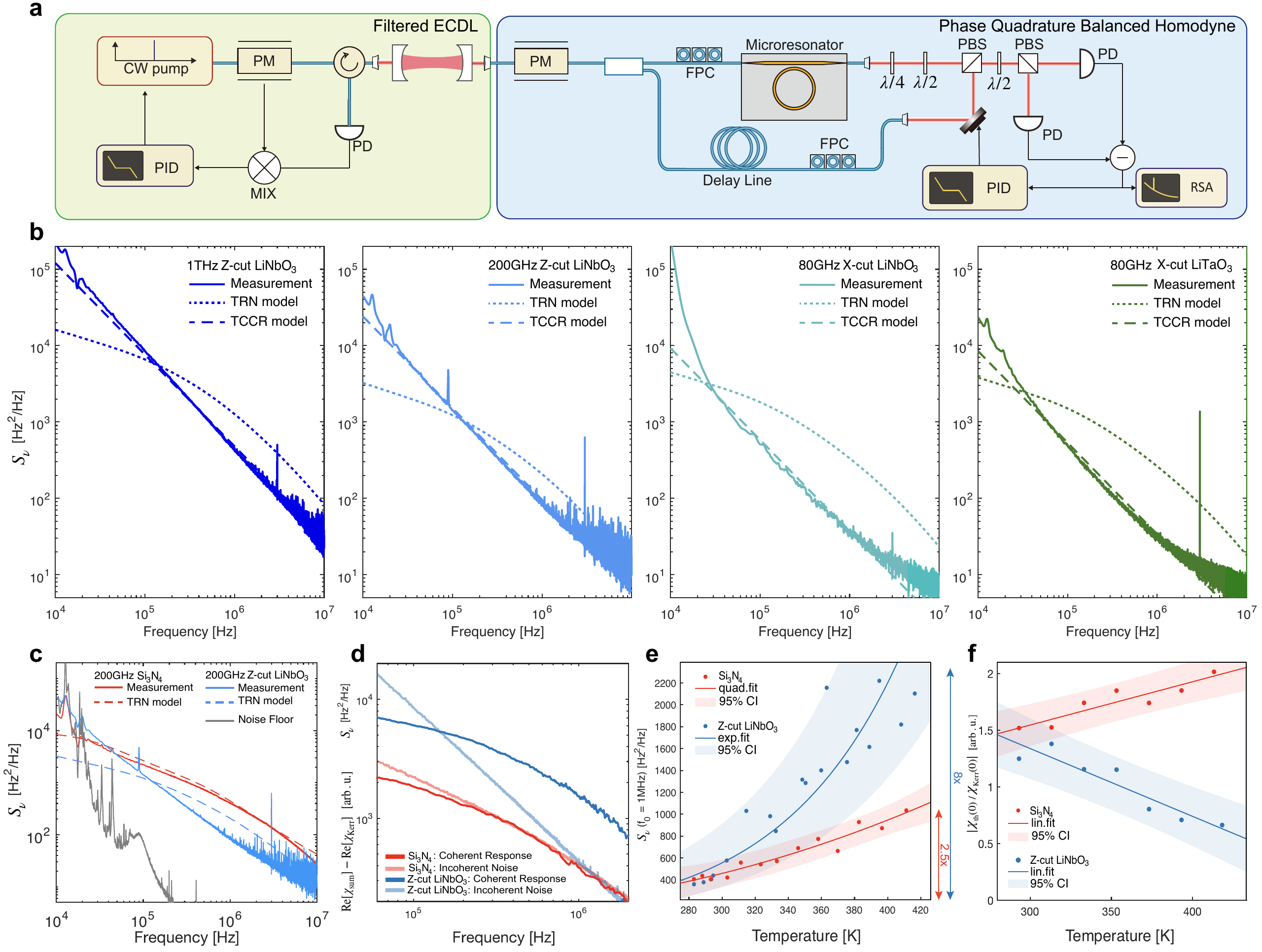}
        %\captionsetup{justification=raggedright,singlelinecheck=false}
        \caption{\textbf{Intrinsic cavity noise in electro-optic materials.}
        (a) Balanced homodyne setup for intrinsic cavity phase noise measurement.
        RSA: real-time electric spectrum analyzer;
        An external cavity diode laser (ECDL) is locked and filtered using the Pound-Drever-Hall (PDH) technique, utilizing a high-finesse Fabry-Perot cavity to reduce frequency noise. 
        The filtered laser is used to drive a balanced Mach-Zehnder interferometer with a cavity on the signal arm. 
        A calibrated phase modulator (PM) is used to generate the calibration tone.
        (b) Single-sided power spectral density (PSD) of intrinsic frequency noise for Z-cut, X-cut \LN and X-cut \LT microresonators with different free-spectral ranges.
        Solid line: measurement results;
        Short dashed line: simulation of thermorefractive noise (TRN).
        Long dashed line: simulation of thermal-charge-carrier-refractive noise (TCCR).
        (c) Single-sided PSD of intrinsic frequency noise for 200~GHz Z-cut \LN and \SiN microresonators. 
        (d) Thermo-optic coherent response function and incoherent cavity noise measured in Z-cut \LN and \SiN microresonators.
        (e)The temperature dependency of the measured cavity noise at an offset frequency of 1~MHz. The scaling behavior of non-ferroelectric Silicon Nitride is fitted with a quadratic function, in accordance with the TRN theory (doubled in the measurement temperature range). The scaling behavior of ferroelectric Lithium Niobate is fitted with an exponential function in accordance with the TCCR theory (eightfold increase). The corresponding 95\% confidence intervals are shaded.
        (f) Temperature scaling of relative strength between photothermal nonlinear and Kerr-nonlinearity: $|\chi_{\mathrm{th}}(0) / \chi_{\mathrm{Kerr}}(0)| $, with linear fit and 95\% confidence intervals.
} 
\label{fig3}       
    }
\end{figure*}

% The observation of pyroelectric-electro-optic effect (PyroEO) counteracting the material's intrinsic thermal-optic effect raises the question of whether the intrinsic cavity noise, previously believed to be dominated by thermorefractive noise (TRN), aligns with the observed reduction. 
We employ a balanced homodyne interferometer (\Fref{fig3}(a)) to probe the intrinsic refractive fluctuations of the optical microresonator.
\Fref{fig3}(b) depicts the intrinsic noise measurement result for Z-cut \LN, X-cut \LN, and X-cut \LT with various Free Spectral Ranges (FSR) compared with the simulation results based on the thermorefractive noise (TRN) model and the thermal-charge-carrier-refractive noise (TCCR) model.
The measurement range is limited by technical noise at low offset frequencies and by shot noise at high offset frequencies (see supplementary material).
\Fref{fig3}(c) compares the intrinsic noise spectra in Z-cut \LN and \SiN microresonators with 200~GHz FSR.
Indeed, we find that the measured cavity frequency noise at high offset frequencies beyond 30~kHz in Z-cut, X-cut \LN and X-cut \LT~ are all significantly lower than in \SiN-based microresonators of the same size and the prediction from thermorefractive noise theory\cite{huangThermorefractiveNoiseSiliconnitride2019,gorodetskyFundamentalThermalFluctuations2004,LEVIN20081941} due to the pyroelectric-electro-optic (PyroEO) effect.
However, we also find large excess noise at low offset frequency , and a distinctive Flicker-like frequency dependency: $S_{\nu}(f)\propto f^{-1.2}$ across a large frequency range from 10~kHz to 10~MHz.
%JZ: low frequency can't be laser noise, since laser is FP filtered.
The difference in frequency scaling of the spectra indicating a different underlying diffusion mechanism behind the noise with lower natural characteristic frequency\cite{vossFlickerNoiseEquilibrium1976}. 

We attribute this to a new and non-thermorefractive noise channel specific to electro-optic materials: the thermal-charge-carrier-refractive noise, analogous to the charge-carrier induced Flicker noise that is commonly observed in semiconductor devices due to charge diffusion and surface trapping effect with a frequency scaling $S_{\nu}\propto f^{-\eta}$ and factor $\eta$ between 0.5 and 1.5 \cite{hungUnifiedModelFlicker1990, jimminchangFlickerNoiseCMOS1994,jayaramanNoiseTechniqueExtract1989}. 
The Brownian motion of charge carriers gives rise to fluctuations in the charge carrier density that induce microscopic electric fields, which in electrical circuits is referred to as Johnson-Nyquist noise \cite{johnsonThermalAgitationElectricity1928,nyquistThermalAgitationElectric1928}. 

Compared to earlier theoretical predictions of charge carrier noise in high-purity silicon used for mirrors in LIGO \cite{brunsThermalChargeCarrier2020,siegelRevisitingThermalCharge2023}, the strong Pockels effect and low mode volume of ferroelectric microresonators entails a dominating TCCR noise, which can be approximately described by:
\begin{equation*}\tag{2}
    S_\mathrm{\nu,TCCR}(\omega,T) 
    = \frac{n^4 r^2 \nu^2 k_B T}{V_{\mathrm{eff}}}{\frac{\sigma(\omega,T)}{\epsilon_0^2 \epsilon_r^2 \omega^2}}
    \label{TCCR}
\end{equation*}
wherein $\omega\gg 1/\tau_d$, with the dielectric relaxation time $\tau_d = \epsilon_r\epsilon_0 / \sigma(0,T)$. 
$\tau_d>1\mathrm{s}$ in \LN~due to the low conductivity.
Here, $n$ is the refractive index, $r$ is the Pockels coefficient for the specific polarization, $k_B$ is the Boltzmann constant, $T$ is the absolute temperature, $V_\mathrm{eff}$ is the effective optical mode volume, $\nu$ is the optical frequency, $\epsilon_0$ is the vacuum permittivity, $\epsilon_r$ is the relative permittivity and $\sigma(\omega,T)$ is the temperature- and frequency-dependent conductivity which is proportional to the charge-carrier diffusion coefficient. 
A detailed derivation of the noise power spectral density and microscopic theory of charge carrier density fluctuations is presented in the supplementary material.
In semiconductors or insulators, due to quantum tunneling and barrier hopping, the conductivity has a frequency dependency $\sigma(\omega,T)\propto \omega^s$, with $s\lesssim 1$ \cite{elliottTheoryConductionChalcogenide1977,mott1971electronic}.
Equation \ref{TCCR} predicts a frequency dependency of the TCCR of $S_{\nu}(\omega) \propto \omega^{-(2-s)}$. 
This frequency dependence arises from a diffusion spectrum at frequencies significantly higher than the natural characteristic diffusion frequency, which scales as $\omega^{-2}$ \cite{vossFlickerNoiseEquilibrium1976,siegelRevisitingThermalCharge2023,brunsThermalChargeCarrier2020} and is corrected with the frequency dependency $\omega^s$ of the diffusion constant (conductivity).
Specifically, Ref.~\cite{mansinghACConductivityDielectric1985} reports $s=0.8$ for Lithium Niobate at room temperature, resulting in $S_\mathrm{\nu,TCCR,LN}\propto \omega^{-1.2}$ in excellent agreement with our measurement result, as shown in \Fref{fig3}(b).
It should be noted that due to different doping conditions and fabrication processes, the conductivity $\sigma(\omega,T)$ can vary by several orders of magnitude and we find quantitative agreement between our model and measurement (\Fref{fig3}(b)) assuming a conductivity of $5\times 10^{-9}$~S/m at a frequency of 1~kHz as reported in Ref.\cite{mansinghACConductivityDielectric1985}.
The additional surface trapping effect and defects can also affect the charge-carrier diffusion process~\cite{jayaramanNoiseTechniqueExtract1989}. %JZ: AC also depends largely, and AC is what really matters, so I changed the \simga(0,T) to \sigma(\omega,T)
During the experiment, we ensure that the on-chip laser power is sufficiently low to prevent any unwanted thermal-locking and photorefractive effects. 
Data measured with different on-chip power are available in the supplementary material.

To further confirm that the Flicker-like cavity noise is not of thermal origin, we compare the incoherent cavity noise with the coherently driven response measurement in \Fref{fig3}(d). According to the fluctuation-dissipation theorem \cite{kuboFluctuationdissipationTheorem1966} (detailed explanation available in the supplementary materials), for the thermal process induced cavity noise, the incoherently measured cavity noise should follow the frequency scaling of the real-part of the coherently driven thermo-optic response $\bigchi_{\mathrm{TO}}$, as the frequency scaling is determined by the same heat diffusion process:
\begin{equation*}\tag{3}
    S_{\nu,\mathrm{TRN}}(\omega)\ \propto\  \mathrm{Re}[\bigchi_\mathrm{TO}(\omega)]\ \cancel{\propto}\  S_{\nu,\mathrm{TCCR}}(\omega).
\end{equation*}
\Fref{fig3}(d) plots that the cavity noise from the \SiN~microresonator aligns well with the driven response, except for a minor deviation at low frequencies due to the amplified technical noise in homodyne measurements. 
Conversely, the cavity noise from the ferroelectric Z-cut \LN~microresonator exhibits a Flicker-like frequency scaling, substantially deviating from the driven response, indicating the measured noise is not related to heat diffusion.
As a final test to prove the non-thermal origin of the residual noise, we examine the temperature scaling of the cavity noise and plot the values of the single-sided noise power spectral density at an offset of 1~MHz and temperatures ranging from 10°C to 140°C (\Fref{fig3}(e)). 
The temperature-dependent thermo-optic response is plotted in \Fref{fig3}(f).
For pure thermorefractive noise, we expect a temperature scaling of $S_{\nu}\propto \alpha_{\mathrm{TO}}^2T^2$.
It is evident that the cavity noise of the \SiN~microresonator adheres to the TRN theory.
Because the conductivity of isolators increases exponentially with temperature\cite{mansinghACConductivityDielectric1985}, we expect and observe 
an exponential scaling of the noise amplitude in Z-cut \LN.
% $ , the cavity noise observed in the ferroelectric Z-cut Lithium Niobate scales significantly faster than the prediction of the TRN theory, as the TRN theory predicts the noise to be doubled in the measurement temperature range while we measured an eightfold increase. 
% The corresponding temperature-dependent coefficient of the coherent response is plotted in \Fref{fig3}(g) indicating a reduction in the gross thermo-optic coefficient $|\alpha_\mathrm{TO}|\propto |\bigchi_{\mathrm{th}}(0)/\bigchi_{\mathrm{Kerr}}(0)|$ as the temperature rises in agreement with the PyroEO effect. % "1KHz" is removed, as the usual way is to say (0) (guanhao, loncar, et al)
% Thus, the observed significantly faster scaling of the cavity noise in the Z-cut \LN~which cannot be attributed to the increase in $|\alpha_\mathrm{TO}|$, which is decreasing. 
% This further confirms that the dominating noise source in the ferroelectric microresonators we studied is not thermorefractive noise, but charge-carrier noise.

%%%%%%%%%%%%%%%%%%%%%%%%%%%%%%%%%%%%%%%%%%%%%%%%%%%%%%%%%%%%%%%%%%%%%%%%%%%%%%%%%%%%%%%%%%%%
%%%%%%%%%%%%%%%%%%%%%%%%%%%%%%%%%%%% Self-Injection Locking  %%%%%%%%%%%%%%%%%%%%%%%%%%%%%%%
%%%%%%%%%%%%%%%%%%%%%%%%%%%%%%%%%%%%%%%%%%%%%%%%%%%%%%%%%%%%%%%%%%%%%%%%%%%%%%%%%%%%%%%%%%%%

\section*{Charge Noise Limited Integrated Pockels Laser}
\begin{figure*}[htbp]
    {
        \centering
        \includegraphics[width=1\linewidth]{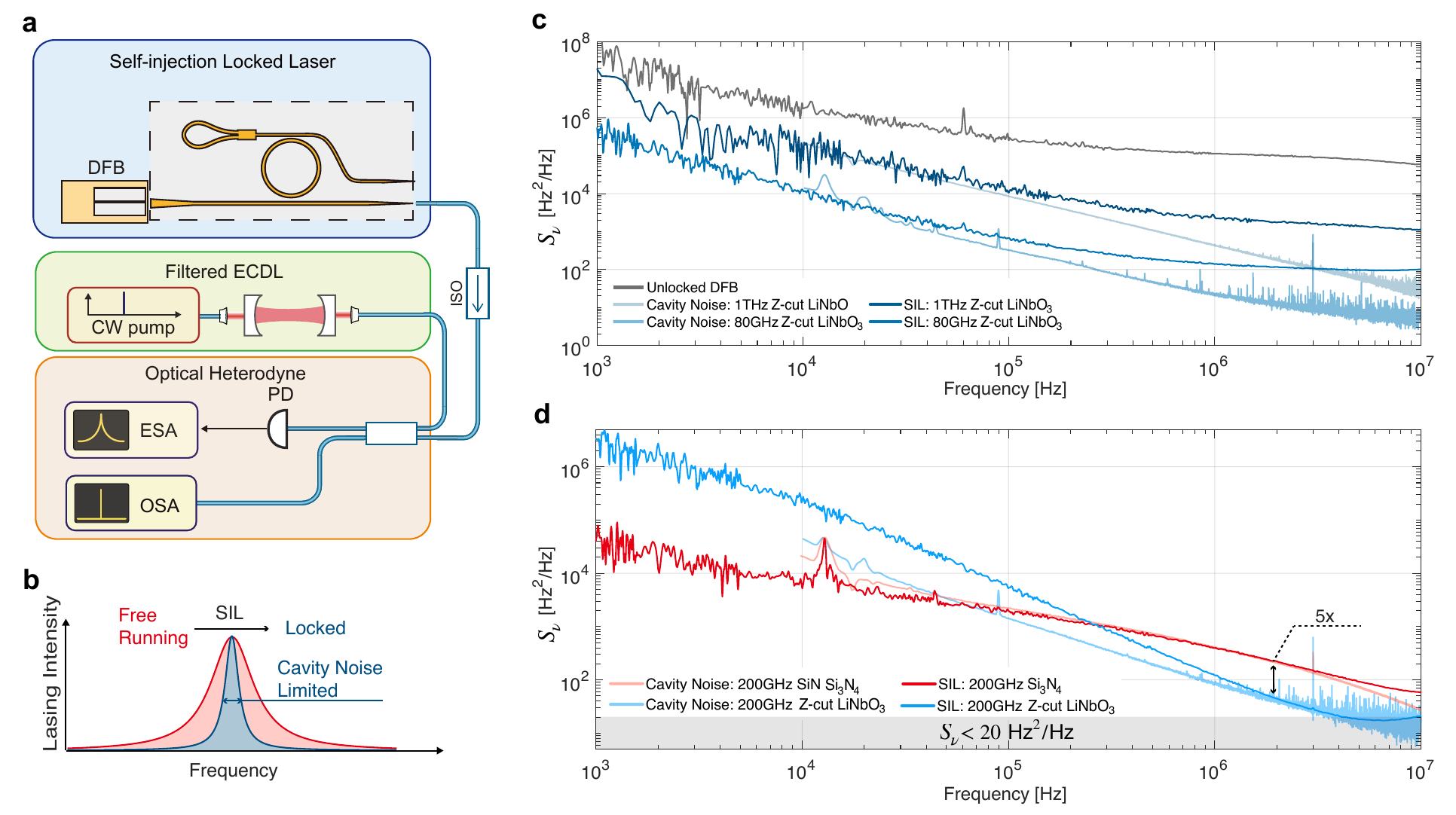}
        \caption{\textbf{Frequency noise of self-injection-locked lasers limited by charge-carrier-refractive noise.}
        (a) Heterodyne-based laser frequency noise measurement setup. 
        ISO: optical isolator;
        ESA: electrical spectral analyzer;
        OSA: optical spectral analyzer;
        The hybrid laser is heterodyne beaten with the filtered external cavity diode laser (the same filtered laser used in the homodyne measurement). The heterodyne signal is detected by a fast photodiode and analyzed by ESA.
        (b) Schematic of frequency reduction of self-injection locking (SIL). Self-injection locking reduces the frequency noise of the laser diode, the optimal performance of the SIL laser is limited by the fundamental cavity noise.
        (c) Singled-sided power-spectral density (PSD) of self-injection locked (SIL) laser (and free-running DFB) frequency noise and intrinsic cavity noise in Z-cut \LN~with FSRs: 1THz, 80GHz.
        (d) Comparison of the intrinsic noise limit and the measured frequency noise of self-injection locked lasing in 200GHz rings made of ferroelectric Z-cut Lithium Niobate and non-ferroelectric Silicon Nitride.
        } 
    \label{fig4}
    }
    \end{figure*}
To demonstrate the technological importance of our discovery of TCCR in electro-optical photoni integrated circuits, we measured the frequency noise spectra of a self-injection locked lasers (\Fref{fig4}(a)), in which the frequency noise of the free-running diode is suppressed by a factor proportional to the square of the microresonator quality factor ($Q^2$) until it hits the fundamental cavity noise limit (\Fref{fig4}(b)). 
\Fref{fig4}(c) plots the frequency noise spectra for the Z-cut \LN~hybrid laser in free-running and self-injection locked states, along with a comparison to the previously measured intrinsic cavity noise in the homodyne experiment.
The linear characterization data for the resonances used in the self-injection-locked laser experiment is available in the supplementary material.
Across all tested devices with FSRs ranging from 80~GHz to 1~THz, the laser frequency noise power spectral density is in quantitative agreement with the fundamental cavity noise limit at offsets up to 100~kHz. 
% Hence, we confirm that the cavity noise in ferroelectric microresonators is reduced by the PyroEO effect and dominated by TCCR noise.
Additionally, \Fref{fig4}(d) plots a comparison of laser and cavity frequency noise spectra of a \SiN-based and a Z-cut \LN-based hybrid integrated lasers, both with 200~GHz FSR. 
We find a substantial noise reduction of up to a factor of five due to the coherent cancellation of the thermorefractive and PyroEO effects at high-offset frequencies compared to the \SiN-based counterpart, thereby affirming the reduction of thermorefractive noise.
The lowest intrinsic laser linewidth of $\Delta\nu = \pi\times 18\mathrm{Hz} = 56\mathrm{Hz}$ is attained in a Z-cut \LN~microresonator with 200~GHz FSR at an offset frequency of 5~MHz.
% However, the situation is drastically different at low offset frequencies, where the ferroelectric microresonators exhibit a much larger refractive noise than non-electro-optic microresonators due to the charge-carrier noise.
%We attribute the discrepancy between laser noise and the intrinsic noise limit in the low-frequency range of this specific measurement to the mode splitting of the cavity resonance used.	

%%%%%%%%%%%%%%%%%%%%%%%%%%%%%%%%%%%%%%%%%%%%%%%%%%%%%%%%%%%%%%%%%%%%%%%%%%%%%%%%%%%%%%%%%%%%
%%%%%%%%%%%%%%%%%%%%%%%%%%%%%%%%%%%%%%%%%   Conclusions     %%%%%%%%%%%%%%%%%%%%%%%%%%%%%%%%
%%%%%%%%%%%%%%%%%%%%%%%%%%%%%%%%%%%%%%%%%%%%%%%%%%%%%%%%%%%%%%%%%%%%%%%%%%%%%%%%%%%%%%%%%%%%

\section*{Discussion and conclusion}
In summary, we have discovered that the intrinsic cavity noise mechanisms in electro-optic materials can significantly differ from that in non-electro-optic materials. 
We demonstrate that the combination of pyroelectricity and electro-optic response contributes to a negative thermo-optic coefficient in Lithium Niobate and Lithium Tantalate, which not only alters the thermo-response but can also coherently reduces the thermo-refractive noise.
While thermal-charge-carrier-refractive noise was deemed negligible in conventional dielectrics and semiconductors optics in the context of gravitational wave detection \cite{brunsThermalChargeCarrier2020,siegelRevisitingThermalCharge2023}, we find that thermal-charge-carrier-refractive noise, supersedes the reduced thermo-refractive noise and becomes the dominant refractive noise limit in integrated photonic devices.
Equally important, the power spectral density of TCCR in ferroelectric photonics exhibits a very different Flicker-like frequency scaling compared to thermo-refractive noise which saturates at low frequencies. 
Broadly viewed, our measurements in high-Q \LT~and \LN~photonic integrated circuit-based microresonators carry substantial implications in terms of achievable performance and noise limitations for a variety of applications including quantum transduction \cite{sahuQuantumenabledOperationMicrowaveoptical2022}, quantum electro-optic experiments \cite{xuBidirectionalInterconversionMicrowave2021}, $\chi^{(2)}$ squeezing \cite{stokowskiIntegratedQuantumOptical2023,nehraFewcycleVacuumSqueezing2022}, coherent communication, LiDAR \cite{snigirevUltrafastTunableLasers2023} and sensing \cite{caleroUltraWidebandhighSpatial2019}. 
We expect this charge-carrier noise might also be a dominating noise in future low energy silicon\cite{dongLowPpUltralowenergy2009,xuMicrometrescaleSiliconElectrooptic2005} and ferroelectric-based electro-optic integrated photonics \cite{abelLargePockelsEffect2019} as these devices approach the Landauer limit for electro-optical conversion. 
We also highlight the technical relevance of TCCR noise in ferroelectric Pockels photonics already today by demonstrating that the noise performance and frequency drift of low noise self-injection locked Pockels lasers are limited by TCCR, in particular at low offset frequencies.
Lastly, we want to point out that our results also provide a novel way to study charge fluctuations in mesoscopic, such as domain wall switching in ferroelectric materials using the exquisite sensitivity and speed of optical measurements, which may advance precision measurement of charge noise and mesoscopic fluctuations.
	
%%%%%%%%%%%%%%%%%%%%%%%%%%%%%%%%%%%%%%%%%%%%%%%%%%%%%%%%%%%%%%%%%%%%%%%%%%%%%%%%%%%%%%%%%%%%
%%%%%%%%%%%%%%%%%%%%%%%%%%%%%%%%%%%%%%%%%   Methods     %%%%%%%%%%%%%%%%%%%%%%%%%%%%%%%%%%%%
%%%%%%%%%%%%%%%%%%%%%%%%%%%%%%%%%%%%%%%%%%%%%%%%%%%%%%%%%%%%%%%%%%%%%%%%%%%%%%%%%%%%%%%%%%%%

\small
\section*{Methods}

 \textbf{DC-shift and coherently-driven nonlinear response measurement.}
The setup follows the approach introduced in \cite{gaoProbingMaterialAbsorption2022} and its diagram is available in supplementary material.
In the DC-shift measurement, an external cavity diode laser (CW Pump, Toptica 1550) is PDH-locked to the resonance cavity under test and sent to a wavemeter (High-Fines WS6-600, with resolution 20~MHz) to continuously record the cavity resonance frequency.
The laser power is reduced to less than 100$\mathrm{\mu W}$ to avoid unwanted thermal nonlinearity and reduce photorefractive-induced drifting.
The temperature is controlled by a thermoelectric cooler attached to the copper chip holder.
In the coherently-driven nonlinear response measurement, the same pump laser with increased power (mW level, but still low enough to avoid thermal locking) is intensity-modulated by an intensity modulator to drive nonlinearity in the cavity. 
A probe external cavity diode laser is tuned to the side of the resonance.
Note that the pump laser and probe laser were placed to different resonances and a tunable band-pass filter is used to extract the probe signal after cavity for lockin-in detection.
Upon traversing the cavity, the pump laser's intensity modulation generates a resonance shift that is subsequently transduced into transmission modulation for the probe laser on the side of resonance. This modulation is then detected by a fast photodiode (cut-off frequency at 125MHz).
The drive input of the intensity modulator and the photodiode output are connected to a vector network analyzer (VNA) for the detection of the coherent response.

\textbf{Balanced homodyne for intrinsic cavity frequency noise measurement.}
The setup follows the approach introduced in \cite{huangThermorefractiveNoiseSiliconnitride2019}, the simplified diagram is shown in \Fref{fig3}(a) and a complete diagram is shown in the supplementary.
An external cavity diode laser used in the previous experiment is PDH-locked to a high-finesse Fabry-Perot cavity with intrinsic linewidth $\sim 70$~kHz, and the notably low-phase noise filtered transmitted light is used to drive the balanced Mach–Zehnder interferometer. 
The resonance of the Fabry-Perot, under the control of an external piezo drive, is aligned with the resonance of the cavity under test. This alignment allows for maximum efficiency in transducing the cavity frequency noise into a phase fluctuation, detectable by the homodyne method.
Before the interferometer, a phase modulator modulated with a known RF tone to provide an absolute calibration peak for this noise measurement.
A piezo mirror is utilized in the path of the local oscillator to ensure the homodyne is locked at the phase quadrature.
The noise floor is measured by replacing the cavity under-test with a 3m long fiber delay line.

 \textbf{Self-injection locked laser frequency noise measurement.}
A distributed-feedback laser (DFB) is self-injection locked to the microresonator under test and beat with the Fabry-Perot cavity filtered ECDL laser (the same filtering method as in previous homodyne measurement). 
The back-reflection phase of the self-injection locking system is controlled by changing the gap between the DFB laser and chip facet, to attain optimal locking conditions. 
For further integration, the reflection phase can also be tuned by an integrated electro-optic phase shifter.
The heterodyne beat signal is then detected by a photodiode and the resulting RF signal is demodulated around the central beat note frequency to reconstruct the phase and quadrature diagram (IQ-diagram) for phase noise analysis.

%%%%%%%%%%%%%%%%%%%%%%%%%%%%%%%%%%%%%%%%%%%%%%%%%%%%%%%%%%%%%%%%%%%%%%%%%%%%%%%%%%%%%%%%%%%%
%%%%%%%%%%%%%%%%%%%%%%%%%%%%%%%%%%%%%%%%%   Miscellaneous     %%%%%%%%%%%%%%%%%%%%%%%%%%%%%%
%%%%%%%%%%%%%%%%%%%%%%%%%%%%%%%%%%%%%%%%%%%%%%%%%%%%%%%%%%%%%%%%%%%%%%%%%%%%%%%%%%%%%%%%%%%%

\section*{Author Contributions}

J.Z. designed the samples with assistance from J.R. and fabricated the samples with assistance from Z.L.. 
Experiments were carried out by J.Z.,G.L., and G.H.. 
The experimental data was analyzed by J.Z with the assistance of J.R. and G.H..
J.Z. and J.R. developed the theoretical and numerical models. 
T.J.K. supervised the work.

\section*{Funding Information}
This work has received funding from the European Research Council (ERC) under the EU H2020 research and innovation programme, grant agreement No. 835329 (ExCOM-cCEO). This material is based upon work supported by the Air Force Office of Scientific Research under award number FA9550-19-1-0250 and the Swiss National Science Foundation (SNSF) grant no. 216493 (HEROIC). J.R acknowledges support from the SNSF under grant no. 201923 (Ambizione)
\section*{Acknowledgements}

We thank Nils J. Engelsen for the insightful discussion on cavity noise and Amirali Arabmoheghi for assistance with the balanced homodyne interferometer. We thank Anat Siddarth for supplying the reference laser noise data.  
The samples were fabricated in the EPFL Center of MicroNanoTechnology (CMi) and the Institute of Physics (IPHYS) cleanroom.

\section*{Competing interests}
 The authors declare no competing financial interests.

\section*{Data Availability Statement} The code and data used to produce the plots within this work will be released on the repository \textit{Zenodo} upon publication of this preprint.

\bibliography{refs}	
\bibliographystyle{naturemag}
\end{document}

\title{Supplementary Information: Pyroelectrically Modified Thermal Response and Intrinsic Cavity Noise in Ferroelectric Photonics}

\title{}

\author{Junyin Zhang}
\affiliation{Institute of Physics, Swiss Federal Institute of Technology Lausanne (EPFL), CH-1015 Lausanne, Switzerland}
\affiliation{Center of Quantum Science and Engineering (EPFL), CH-1015 Lausanne, Switzerland}

\author{Zihan Li}
\affiliation{Institute of Physics, Swiss Federal Institute of Technology Lausanne (EPFL), CH-1015 Lausanne, Switzerland}
\affiliation{Center of Quantum Science and Engineering (EPFL), CH-1015 Lausanne, Switzerland}

\author{Johann Riemensberger}
\affiliation{Institute of Physics, Swiss Federal Institute of Technology Lausanne (EPFL), CH-1015 Lausanne, Switzerland}
\affiliation{Center of Quantum Science and Engineering (EPFL), CH-1015 Lausanne, Switzerland}

\author{Grigory Lihachev}
\affiliation{Institute of Physics, Swiss Federal Institute of Technology Lausanne (EPFL), CH-1015 Lausanne, Switzerland}
\affiliation{Center of Quantum Science and Engineering (EPFL), CH-1015 Lausanne, Switzerland}

\author{Guanhao Huang}
\affiliation{Institute of Physics, Swiss Federal Institute of Technology Lausanne (EPFL), CH-1015 Lausanne, Switzerland}
\affiliation{Center of Quantum Science and Engineering (EPFL), CH-1015 Lausanne, Switzerland}

\author{Tobias J. Kippenberg}
\email[]{tobias.kippenberg@epfl.ch}
\affiliation{Institute of Physics, Swiss Federal Institute of Technology Lausanne (EPFL), CH-1015 Lausanne, Switzerland}
\affiliation{Center of Quantum Science and Engineering (EPFL), CH-1015 Lausanne, Switzerland}

\maketitle
\tableofcontents

%%%%%%%%%%%%%%%%%%%%%%%%%%%%%%%%%%%%%%%%%%%%%%%%%%%%%%%%%%%%%%%%%%%%%%%%%%%%%%%%%%%%%%%%%%%%
%%%%%%%%%%%%%%% Coherent response and Incoherent cavity Noise %%%%%%%%%%%%%%%%%%%%%
%%%%%%%%%%%%%%%%%%%%%%%%%%%%%%%%%%%%%%%%%%%%%%%%%%%%%%%%%%%%%%%%%%%%%%%%%%%%%%%%%%%%%%%%%%%%
\section{Intrinsic Cavity Noise: thermorefractive noise and Charge Noise}
\noindent The thermorefractive noise arises from the thermo-optic response of optical cavities and is limiting the stability of the resonance frequency of dielectric optical resonators and hybrid integrated lasers. 
It occurs when temperature fluctuations induce corresponding refractive index fluctuations, which can be quantified by the thermo-optic coefficient $\frac{dn}{dT}$. 
These temperature fluctuations stem from the fundamental temperature fluctuation resulting from the heat diffusion process. 
An estimation of the fundamental temperature fluctuation can be obtained using the following equation:
\begin{equation}
        \langle \Delta T \rangle ^2 = \frac{k_B T^2}{C_v},
\end{equation}
wherein $k_B$ denotes the Boltzmann constant, $T$ the absolute temperature, and $C_v$ the heat capacity. 

\subsection{thermorefractive noise calculation based on the fluctuation-dissipation theorem and heat diffusion}
\noindent 
To contrast with the following description of cavity noise that originates from charge fluctuation and  diffusion noise, let's consider the well-established treatment of the thermorefractive noise based on the conventional fluctuation dissipation theorem \cite{levinInternalThermalNoise1998,LEVIN20081941,ThermorefractiveNoiseWhispering}.
The refractive index change due to the scalar temperature fluctuation field $\delta T(\vec{r})$ can be written as:
\begin{equation}
	\frac{\delta n}{n} = -\frac{\int \delta T(\vec{r}) \varepsilon_0 \sqrt{\varepsilon_r(\vec{r})}\beta(\vec{r}) |u(\vec{r})|^2 \ \mathrm{d}^3 r}{\int  \varepsilon_0 \varepsilon_r(\vec{r}) |\vec{u}(\vec{r})|^2\ \mathrm{d}^3 r},
\end{equation}
where $\beta = \frac{dn}{dt}$ is the thermo-optic coefficient. 
The refractive index modulation $\delta n$ can be written in the form:
\begin{equation}
	\delta n = \int \delta T(\vec{r})q(\vec{r})\ \mathrm{d}^3 r, 
\end{equation}
with:
\begin{equation}
	q(\vec{r}) = -\frac{ n(\vec{r}) \varepsilon_0 \sqrt{\varepsilon_r(\vec{r})}\beta(\vec{r}) |u(\vec{r})|^2 }{\int \varepsilon_0 \varepsilon_r(\vec{r}) |\vec{u}(\vec{r})|^2\ \mathrm{d}^3 r}.
\end{equation}
It has been proven\cite{levinInternalThermalNoise1998} that the noise spectra of $\delta n$ can be obtained by
\begin{equation}
	S_{\delta n}(f) = \frac{2 k_B T}{\pi^2 f^2}\frac{W_\mathrm{diss}}{F_0^2},
\end{equation}
where the dissipated power $W_\mathrm{diss}$ is calculated by injecting the system with an entropy density with a period of $1/f$ ($\omega = 2 \pi f$) and a spatial profile that matches the temperature fluctuation profile driven by the optical mode $q(\vec{r})$
\begin{equation}
	\frac{\delta s(\vec{r})}{\mathrm{d}r^3} = F_0 \cos(\omega t)q(\vec{r}).
\end{equation}
The fluctuation-dissipation theorem-based method proves to be a practical approach as it can be readily simulated using the finite-element method. 
In the case of simulating thermorefractive noise in COMSOL \cite{kondratievThermorefractiveNoiseWhispering2018}, the procedure involves periodic injection of a heat source with spatial dependency $q(\vec{r})$. 
By introducing this heat source, a temperature field $\delta T$ can be obtained, enabling the calculation of the dissipated energy $W_\mathrm{diss}$ using the following equation: 
\begin{eqnarray}
	W_\mathrm{diss}(\omega) = \int{\frac{\pi\kappa}{\omega T}(\nabla \delta \tilde{T})^2\  \mathrm{d}^3 r},
\end{eqnarray} 
where $\delta\tilde{T}$ is the time-domain Fourier transform of the time-dependent temperature field $\delta T$ induced by the periodically injected heat-source $q(\vec{r})$, the $\kappa$ is the heat conductivity of the material, $T$ refers to the environment temperature.
\begin{figure*}[htb]
	{
		\centering
		\includegraphics[width=1.0\linewidth]{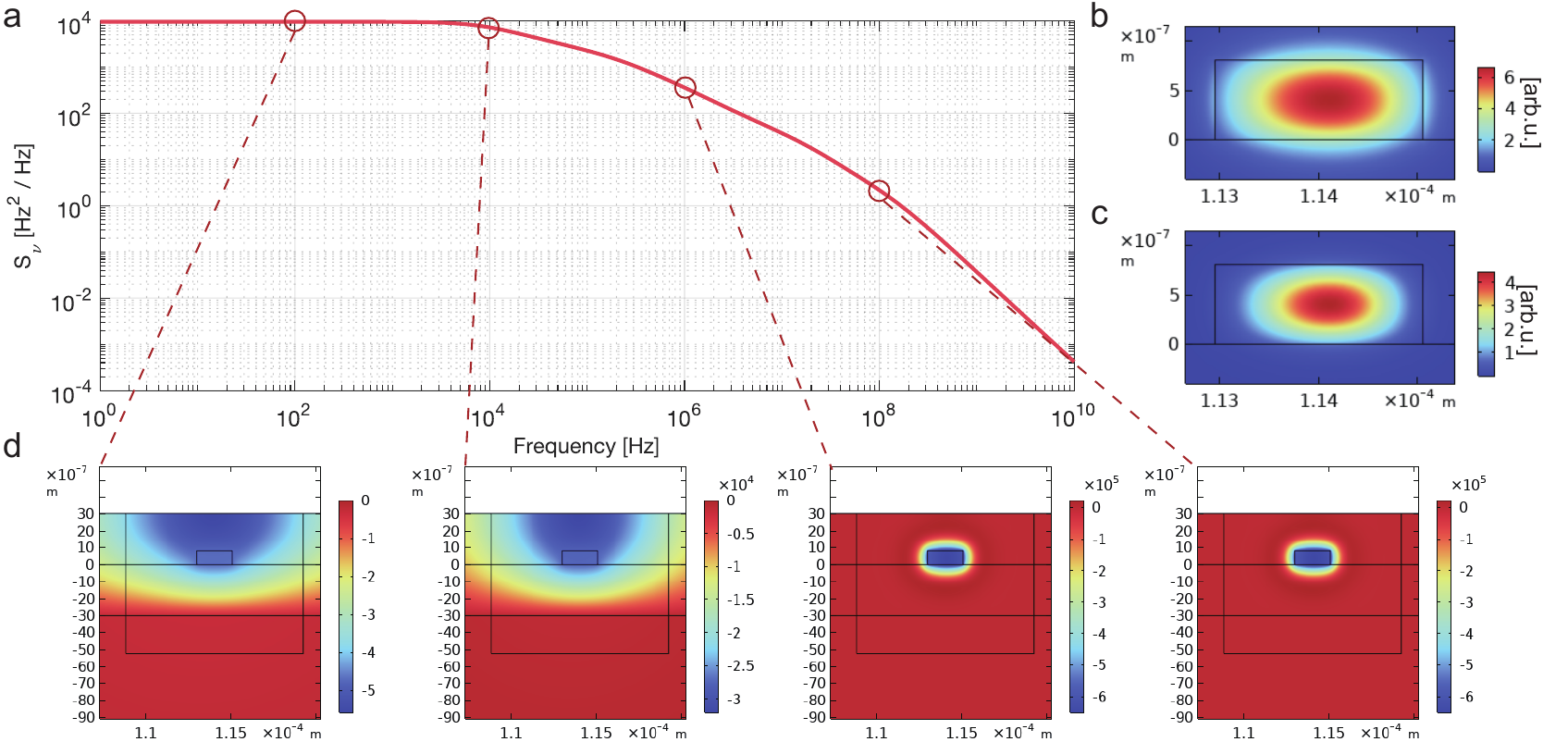}
		\caption{\textbf{Thermorefractive noise simulation for 200GHz \SiN microresonator and the cross-section of dissipated temperature field $\delta T$.}
			The upper panel (a) displays the thermorefractive noise simulation for in a 200~GHz \SiN-based optical microresonator using the fluctuation-dissipation theorem based finite-element method. (b) The optic mode profile. (c) The driving heat-source term corresponding to $q(\vec{r})$ introduced in the fluctuation-dissipation calculation.  In the lower panel (d), the optical mode profile and the dissipation field $\delta T$ are presented for different modulation frequencies. 
		} 
		\label{TRN_crosssection}       
	}
\end{figure*}

\subsection{Order estimation of the charge carrier noise in Lithium Niobate from Johnson Noise}
\noindent In contrast to dielectrics like \SiO and \SiN, we have observed the existence of a new noise channel in  ferroelectric materials with electric-optic response such as \LN and \LT that is not thermally driven and does not match the coherent thermal resonse measurement nor the simulations of the frequency dependence of the thermal dissipation.
To resolve this conundrum, we propose a new model for the Flicker-like cavity noise of ferroelectric optical microresonators that is not driven by microscopic temperature fluctuations but by fluctutations of thermally excited charge carriers that induce an electric field that modulates the cavity frequency through the Pockels effect $r$. 
This phenomenon is well-known in the field of semi-conductor based electronics as the Flicker-noise that is found in field-effect transistors \cite{vossFlickerNoiseEquilibrium1976} and its optical counterpart thermo-charge-carrier-rerfractive (TCCR) noise has also been recently discussed in the context of gravitational wave detection \cite{brunsThermalChargeCarrier2020,siegelRevisitingThermalCharge2023}.

First, we want to present a simple model of the noise that is induced by charge carrier fluctuations and the Pockels effect in ferroelectric materials based on an equivalent circuit model of the z-cut \LN thin film that we can describe as an RC-element with electrodes at the top and bottom surface of the \LN film.
In a resistor with resistance $R$ at an absolute temperature $T$, the Brownian motion of charge carriers, such as electrons, results in a white voltage noise across the resistor. This phenomenon is commonly known as Johnson Noise with a white power spectral density of:
\begin{equation}
    S_{U}(f) = 4 k_B T R,
\end{equation}
where $k_B$ is the Boltzmann-constant. 
In the following, we will demonstrate that the relatively weak voltage fluctuation in \LN-based microresonators induces a measurable cavity frequency noise $S_\nu(f)$, which can exceed the level of thermorefractive noise.
For the sake of simplicity, we will consider a rectangular waveguide etched from a Z-cut \LN film, with a width of $w$ and a height of $h$ and restrict our analysis to the strongest Pockels tensor component acting on the fundamental TE mode $r_{13}$.
The estimation for x-cut \LN, \LT, and similar materials may be conducted in a similar manner.
Considering the electro-optic response originating from the electric field in the Z-direction (indicated by the Pockels coefficient $r_{13} = 10  \mathrm{pm/V}$), the resulting refractive index change due to this electric field in the Z-direction is given by:
\begin{equation}
    \delta n = \frac{1}{2}n^3 r_{13} E_{z},
\end{equation}
where $n$ represents the refractive index and $E_z$ represents the Z-component of the electric field. As a result, the cavity resonances will experience a shift:
\begin{equation}
    \delta f = f_0 \frac{\delta n}{n},
\end{equation}
where $f_0$ is the optical frequency.

\begin{figure*}[htb]
    {
        \centering
        \includegraphics[width=1\linewidth]{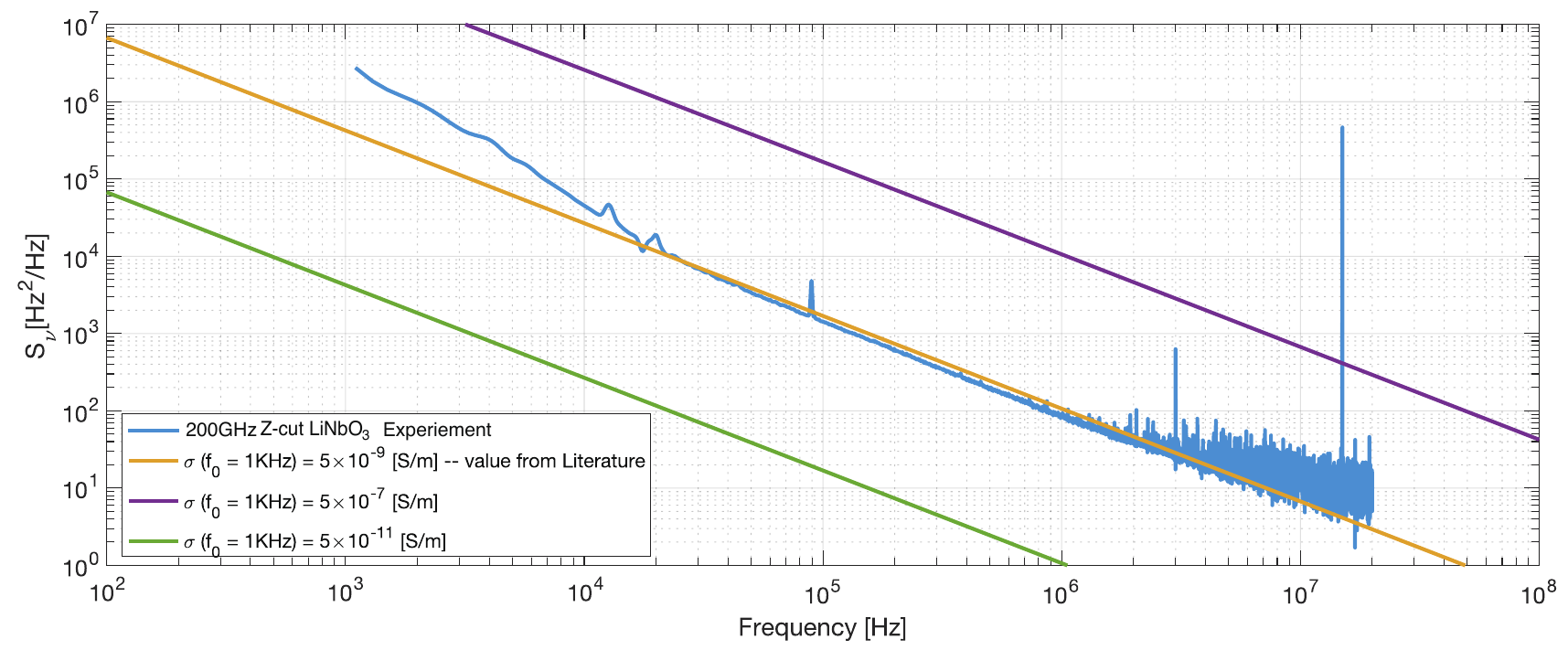}
        \caption{\textbf{Charge-carrier noise estimation for 200GHz Z-cut Lithium Niobate using the simple Johnson Noise model.} 
        The Johnson Noise model is utilized to simulate the charge-carrier noise in a z-cut Lithium Niobate microresonator with a free spectral range (FSR) of 200 GHz. The conductivity values considered for the calculations are: $\sigma = 5\times 10^{-9},\ 5\times 10^{-7},\ 5\times 10^{-11}\ [S/m]$ at offset frequency $f_0=1kHz$, and follows the frequency dependency $\sigma(f)\propto f^{s},\ s=0.8$. Among which, $\sigma = 5\times 10^{-9} [S/m]$ and $s=0.8$ is the value measured in bulk \LN in ref.\cite{mansinghACConductivityDielectric1985}. Additionally, the measured cavity noise in the 200 GHz z-cut Lithium Niobate microresonator is provided for comparison.
        } 
\label{JohnsonNoise}       
    }
\end{figure*}

Considering a micro-resonator with a length of $L$, the resistance between the top and bottom of the resonator can be expressed as:
\begin{equation}
    R = \frac{h}{\sigma w L},
\end{equation}
where $\sigma\approx 1\times 10^{-10}\ [\mathrm{S}/\mathrm{m}]$ is the conductivity for bulk \LN \cite{lucasHightemperatureElectricalConductivity2022}. 
For a 200~GHz microring resonator, we approximate the resistance as $R \approx 4.9 \mathrm{T\Omega}$. 
However, it is important to note that the actual value may be lower due to surface conductance, in particular in x-cut \LN-based microresonators. 
The corresponding Johnson Voltage noise power spectral density can then be calculated as:
\begin{equation}
    S_{U}(f) = 4 k_B T \frac{h}{\sigma w L},
\end{equation}
which generates an electric field noise:
\begin{equation}
    S_{E}(f) = \frac{S_{U}(f)}{h^2}.
\end{equation}
This electric field noise power spectral density $S_{E}(f)$ will directly lead to the cavity frequency noise $S_{\nu}(f)$ via the Pockels effect:
\begin{align}
    S_{\nu}(f) &= \left(\frac{1}{2}n^2f_0 r_{33}\right)^2 S_{E}(f)\\
    &= \dfrac{n^4 r_{33}^2 k_B T}{\sigma V}f_0^2
\end{align}
where $V = wLh$ is approximately the mode-volume of the optical field inside the resonator. 
Consider a 200~GHz resonator with radius $r = 114\,\mu$m and length $L = 2\pi r$, with height $h = 0.7\,\mu$m and width $w = 2\,\mu$m, the cavity frequency noise induced by Johnson noise can be calculated as follows:
\begin{equation}
    S_{\nu}(f) = 2.26\times 10^{13}[Hz^2/Hz] \gg S_{\nu, \mathrm{TRN}}(f).
\end{equation}
Hence, we find that the Johnson noise-induced cavity frequency noise can have a notable impact on the ferroelectric photonics devices under consideration. 
% However, it is important to note a limitation in the previous derivation. 
The cavity noise is not white but exhibits a pink spectrum due to the \LN film not only acting as a resistor, but also as a capacitor with a higk-K dielectric filling with $\varepsilon_r=27.9$.
Treating the \LN-based waveguide ring as a simple plate capacitor and ignoring the fringe fields, we can write the capacitance as:
\begin{equation}
	C = \varepsilon_0 \varepsilon_r \frac{wL}{h}.
\end{equation}
For our 200~GHz FSR microresonator we find $C\approx 0.5 \mathrm{pF}$. 
The combination of the capacitor $C$ and resistance $R$ introduces a low-pass filter effect to the Johnson noise, leading to the following:
\begin{equation}
    S_{U}(f) = \frac{4 k_B R T}{1 + (2\pi R C f)^2} 
\end{equation}
with $\varepsilon_0$ to be the vacuum permittivity and $\varepsilon_r$ is the relative permittivity. 
The 3~dB cutoff frequency of the low-pass filter of the 200~GHz FSR z-cut \LN based microresonator follows as:
\begin{equation}
    f_\mathrm{3dB} = \frac{1}{\tau_{RC}} = \frac{1}{2\pi RC} =  0.4\,\mathrm{Hz}. 
\end{equation}
The time-constant $\tau_{RC}$ is usually referred as the \textit{dielectric relaxation time} and only depends on the material properties $\epsilon_r$ and $\sigma$ but not the geometry:
\begin{equation}
   \tau_{RC} = 2\pi RC = 2\pi\frac{\varepsilon_0\varepsilon_r}{\sigma}.
\end{equation}
The cavity noise power spectral density that is induced by the transduction of Johnson voltage noise via the Pockels effect follows at last: 
\begin{equation}
    S_{\nu}(f) = \frac{n^4 r_{33}^2 k_B T}{\sigma(f) V}{\frac{1}{1+(2\pi\frac{\varepsilon_0\varepsilon_r}{\sigma(f)})^2}}f_0^2
    \label{TCCRJohnson}
\end{equation}
Considering that the material conductivity can vary by several orders due to different fabrication processes and environmental factors, as well as being significantly affected by surface effects, we present an estimation of the Johnson noise-induced cavity noise using multiple conductivity values $\sigma$. 
The obtained results, along with the experimentally measured cavity noise, are plotted in \fref{JohnsonNoise}.
The obtained results unequivocally demonstrate that the inclusion of charge carrier noise is essential when analyzing cavity noise in ferroelectric photonic components. 
While the dielectric constants of \LN and \LT are well known, we find that the reported values of the optical conductivity for these materials vary greatly with stochiometric composition, doping, and depending on the crystal growth or additional treatment steps such as ion slicing with literature supporting values including $5\times 10^{-9}[\mathrm{S/m}]$\cite{mansinghACConductivityDielectric1985} at room temperature at modulation frequency $f_0 = 1\mathrm{kHz}$ and $2\times 10^{-11}[\mathrm{S/m}]$\cite{lucasHightemperatureElectricalConductivity2022}at room temperature at modulation frequency $f_0 = 0.42\mathrm{kHz}$.

Nevertheless, it is important to note that the estimation based on a simple equivalent circuit model can give a reasonable approximation of the noise for realistic values of the optical conductivity $\sigma$. 
Moreover, both from the established theory of the conductivity \cite{anderson1975model} and experimental measurements of \LN crystals \cite{mansingh1985ac}, we find that  $\sigma$ in \LN increases strongly with frequency according to a $f^{0.8}$ power law at room temperature. 
Inserting the frequency dependence of $\sigma$ into equation \ref{TCCRJohnson}, we obtain excellent agreement with the measured power law dependence of Flicker noise $f^{1.2}$ in the z-cut \LN based optical microring resonator.
In the subsequent section, we will introduce a comprehensive microscopic model that utilizes the charge-diffusion model with fluctuation dissipation theorem to accurately calculate the cavity frequency noise induced by charge carriers.

%%%%%%%%%%%%%%%%%%%%%%%%%%%%%%%%%%%%%%%%%%%%%%%%%%%%%%%%%%%%%%%%%%%%%
\subsection{Charge-carrier-refractive noise calculation based on the fluctuation-dissipation theorem and charge diffusion}
In a similar manner to thermorefractive noise, where temperature fluctuations are associated with the diffusion process of the heat, charge-carrier-refractive noise involves charge fluctuations linked to the diffusion process of the charge density. 
This relationship can be expressed as formulated in \cite{brunsThermalChargeCarrier2020}:
\begin{equation}
    \frac{\partial \eta}{\partial t} = D\left( \nabla^2 \eta - \frac{1}{l_D^2}\eta \right) + F(\vec{r},t)
\end{equation}
where $\eta = \rho_{cc} - \rho_0$ is the deviation of charge density, corresponding to $\delta T$ in the derivation for thermorefractive noise, with $\rho_0$ denoting the material charge-carrier density at steady state and $\rho_{cc}$ denoting the charge density in the presence of charge fluctuation. 
$D$ is the diffusion constant for the charge density, which is related to the charge carrier mobility and the conductivity of the material. 
The Langevin force term $F(\vec{r},t)$ is the current flow density that drives the charge density fluctuation, which corresponds to Johnson Noise when averaged over the volume of the \LN-based optical microresonator. 
$l_D$ is the Debye length for the charge screening effect in the dielectric:
\begin{equation}
    l_D = \sqrt{\frac{\varepsilon_0 \varepsilon_r k_B T}{n_0 e^2}}
\end{equation}
This charge fluctuation would induced a electric field $\vec{E} = E_3\vec{e_3}$, where we again only treat the electric field fluctuation that drives the strongest coefficient of the Pockels effect for the TE fundamental mode in z-cut \LN:
% For simplicity, we first consider the contribution of $E_3 r_{33}$, which would leads to the refractive index change due to electro-optics interation:
\begin{equation}
    \frac{\delta n}{n} = -\frac{\int E_3(\vec{r})r_{13}(\vec{r}) n^2 \varepsilon_0 \sqrt{\varepsilon_r(\vec{r})} |u(\vec{r})|^2 /2\ \mathrm{d}^3 r}{\int  \varepsilon_0 \varepsilon_r(\vec{r}) |\vec{u}(\vec{r})|^2\ \mathrm{d}^3 r}
\end{equation}
the charge density induced electric field $E_3$ is related to the charge density field $\eta(\vec{r})$ by:
\begin{equation}
    \nabla\cdot\vec{E} = \frac{\eta}{\varepsilon_0}
\end{equation}
The equation above cannot be directly rewritten in a form that involves only local interaction ($r\to r$) within the integral $\delta n = \int \eta(\vec{r})q(\vec{r})\ \mathrm{d}^3 r$, where $\eta(\vec{r})$ follows a diffusion equation. 
Instead, to calculate noise power spectral density using the fluctuation-dissipation method, we need to solve for the electric field that is induced as a consequence of the charge carrier fluctuation. 
For this, we can  method.

Since the Maxwell equation is linear, we can write down the solution of the electric field $\vec{E}$ and in particular $E_3$ formally as: 
\begin{equation}
    E_3 = \vec{e_3}\cdot \nabla^{-1} \eta(\vec{r})/\varepsilon_0.
    \label{E3_formal}
\end{equation}
A \textit{Green function} $G(\vec{r},\vec{r^\prime})$ always exists for this linear mapping, such that: 
\begin{equation}
    E_3(\vec{r}) = \int G(\vec{r},\vec{r^\prime})\eta(\vec{r^\prime})\ \mathrm{d}^3 r.
\end{equation}
Hence, the fractional-refractive index variation can be written as:
\begin{equation}
    \frac{\delta n}{n}  = \int q(\vec{r}) G(\vec{r},\vec{r^\prime})\eta(\vec{r^\prime})\ \mathrm{d}^3 r\ \mathrm{d}^3 r^\prime,
\end{equation}
with
\begin{equation}
    q(\vec{r}) = -\frac{ r_{13} n^2 \varepsilon_0 \sqrt{\varepsilon_r} |u(\vec{r})|^2 /2\ }{\int  \varepsilon_0 \varepsilon_r |\vec{u}(\vec{r})|^2\ \mathrm{d}^3 r}
\end{equation}
Utilizing the symmetry of the Green function with respect to the spatial coordinates $G(\vec{r},\vec{r^\prime}) = G(\vec{r^\prime},\vec{r}).$, we find:
\begin{equation}
\begin{aligned}
   \frac{\delta n}{n}  &= \int q(\vec{r^\prime}) G(\vec{r^\prime},\vec{r})\eta(\vec{r})\ \mathrm{d}^3 r\ \mathrm{d}^3 r^\prime\\
    &= \int \eta(\vec{r}) \left(\int q(\vec{r^\prime}) G(\vec{r^\prime},\vec{r})\ \mathrm{d}^3 r^\prime \right) \mathrm{d}^3 r \\
    &= \int \eta(\vec{r}) Q(\vec{r}) \mathrm{d}^3 r.
\end{aligned}    
\label{tccr_long}
\end{equation}
With the form-factor $Q(\vec{r})$ defined by:
\begin{equation}
	Q(\vec{r}) = \int q(\vec{r^\prime}) G(\vec{r^\prime},\vec{r})\ \mathrm{d}^3 r^\prime,
	\label{Qr}.
\end{equation} 
The noise spectra can be derived using the method outlined in Ref.\cite{siegelRevisitingThermalCharge2023}, applying the fluctuation-dissipation theorem. Here,  a generalized periodic driving force $\vec{F}=1[J]\times \nabla Q(\vec{r})e^{i\omega t}$ is applied and the charge-carrier diffusion equation can be written as:
\begin{equation}
    (-\frac{\partial}{\partial t} + D\nabla^2 - \frac{\mu \rho_0}{\varepsilon_0\varepsilon_r})\delta\rho = \rho_0 \mu \nabla\cdot \vec{F},
\end{equation}
where $D = \frac{\mu}{e}k_B T$ is the diffusion constant, $e$ is the electric charge of the carrier, $\rho_0$ is the background charge density, and $\mu$ is the generalized mobility, the frequency-dependent conductivity $\sigma(\omega)$ (AC-conductivity) can be introduced and the corresponding relaxation time $\tau_D(\omega)$ can be introduced as:
\begin{equation}
    \frac{1}{\tau_D(\omega)} = \frac{\sigma(\omega)}{\varepsilon_0\varepsilon_r} = \frac{\mu(\omega) \rho_0}{\varepsilon_0\varepsilon_r}.
\end{equation}
Solving the above diffusion equation allows us to determine the charge density variation $\delta \rho$ and the corresponding current density $\vec{J}$. The instantaneously dissipated power is given by:
\begin{equation}
    W_{\mathrm{inst}} = \int d^3 r \vec{F}\cdot\vec{J},
\end{equation}
The time-averaged value of this, the dissipated power $W_{\mathrm{diss}}$, is directly linked to the noise spectra density, which is expressed as
\begin{equation}
    S_{\delta n/n}  = \frac{8k_B T}{\omega^2}\frac{W_{\mathrm{diss}}}{1[J]^2}.
\end{equation}
This is further related to the frequency noise spectra through the relation:
\begin{equation}
    S_{\nu,\mathrm{TCCR}} = \nu^2 S_{\delta n/n} ,
\end{equation}
where $\nu$ is the optical frequency.

In the simulation software \textit{COMSOL} (simulation codes available in \textit{Zenodo} upon publication), $Q(\vec{r})$, we use the electro-static simulation noise to solve Poisson's equation for the "charge density" $\delta q(\vec{r})$. We obtain hence a formal solution to equation \ref{Qr} by retrieving the numerical result for the corresponding "electric field" $E_\mathrm{3,calc}$, which we insert into equation \ref{tccr_long} and proceed with the standard formalism for the calculation of the noise power spectral density using the fluctuation dissipation theorem.
\begin{figure*}[htb]
    {
        \centering
        \includegraphics[width=\linewidth]{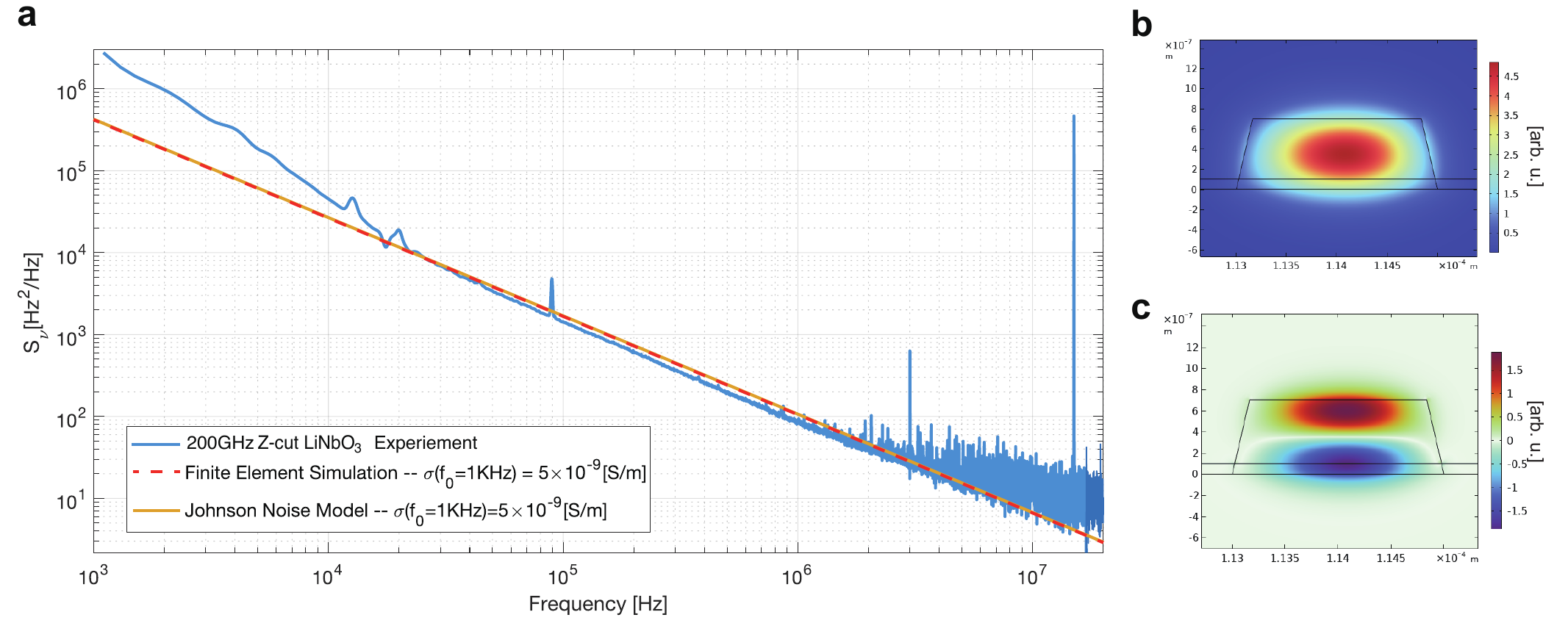}
        \caption{\textbf{Charge-carrier noise estimation for 200GHz Z-cut Lithium Niobate based on Finite-element Method.}
        (a) The measured cavity noise in 200GHz Z-cut \LN microresonator, the simulation result obtained from finite-element method (COMSOL) based charge-diffusion model and the analytic result from Johnson Model. (b) Crossection of the optic mode profile. (c) The driving dissipation term $Q(\vec{r})$.
        } 
\label{JohnsonNoise_COMSOL}       
    }
\end{figure*}

%%%%%%%%%%%%%%%%%%%%%%%%%%%%%%%%%%%%%%%%%%%%%%%%%%%%%%%%%%%%%%%%%%%%%%%%%%%%%%%%%%%%%%%%%%%%
%%%%%%%%%%%%%%%%%%%%%%%%%%%%%%%%%%%% Derivation PyroEO %%%%%%%%%%%%%%%%%%%%%%%%%%%%%%%%%%%%
%%%%%%%%%%%%%%%%%%%%%%%%%%%%%%%%%%%%%%%%%%%%%%%%%%%%%%%%%%%%%%%%%%%%%%%%%%%%%%%%%%%%%%%%%%%%
\section{Derivation of the PyroEO Effective Thermo-optic coefficient}
\noindent In dielectric materials, changes in temperature result in variations of the refractive index, which is known as the thermo-optic effect. 
In ferroelectric materials exhibiting both electro-optic response and pyroelectric effect, temperature fluctuations induce a pyroelectric charge that further modifies the refractive index through the electro-optic effect. 
We refer to this combined pyroelectric effect and electro-optic effect as \textit{PyroEO}. 
This effect has also been investigated in bulk \LN waveguides fabricated by titanium in-diffusion \cite{pointelLithiumNiobateOptical2020}.
Here, we present a preliminary estimation of the PyroEO effect, while detailed calculations can be performed using the finite-element method (relevant codes will be made available on \textit{Zenodo} upon publication). 
We initially focus on the PyroEO effect affecting TE-modes in Z-cut Lithium Niobate. 
Through the pyroelectric effect, a temperature change $\Delta T$ induces a polarization density $\Delta P$ along the Z-direction as follows:
\begin{eqnarray}
	\Delta \vec{P} = p \Delta T\  \vec{e_z}.
\end{eqnarray}
In \LN, the pyroelectric constant $p$ has a value of $-8.3 \times 10^{-5} [C/(\mathrm{K\cdot·m})]$. This constant signifies that when the temperature of the nanowaveguide increases, a spontaneous polarization is generated along the -Z direction. 
This spontaneous polarization gives rise to an electric field within the dielectric which is opposite to the polarization direction. Consequently, the pyroelectric charges result in the induction of a pyroelectric field:
\begin{eqnarray}
	\Delta \vec{E} = -\frac{\Delta \vec{P}}{\varepsilon_0 \varepsilon_r} = -\frac{p}{\varepsilon_0 \varepsilon_r} \Delta T.
\end{eqnarray}
The electric field induced by the pyroelectric charges leads to a change in the refractive index through the Pockels effect. This change is given by the equation:
\begin{equation}
	\begin{aligned}
		\Delta n &= -\frac{1}{2}n^3 r_{13}\Delta \vec{E} \cdot \vec{e_z}\\ 
		&= \frac{n^3 r_{13}p}{2\varepsilon_0 \varepsilon_r}\Delta T.
	\end{aligned}
\end{equation}
Here, $n = 2.21$ denotes the refractive index for o-light in \LN, specifically corresponding to the TE-mode in a z-cut \LN-based optical waveguide. The strongest Pockels coefficient acting on the light field is $r_{13} = 9.6\ \mathrm{pm/V}$. 
Consequently, the PyroEO effect gives rise to an negative effective thermo-optic coefficient (TOC):
\begin{equation}
	\dfrac{\partial n}{\partial T}\bigg\vert_\mathrm{PyroEO} = 
	\frac{n^3 r_{13} p }{2\varepsilon_0 \varepsilon_r} = 
	-1.74\times 10^{-5}\ [\mathrm{1/K}].
\end{equation}
Similarly, for the TE-mode propagating in the x-cut \LN-based optical waveguide, we can calculate the effective TOC induced by the PyroEO effect using the same method replacing $r_{13}$ with the tensor component $r_{33}$.

%%%%%%%%%%%%%%%%%%%%%%%%%%%%%%%%%%%%%%%%%%%%%%%%%%%%%%%%%%%%%%%%%%%%%%%%%%%%%%%%%%%%%%%%%%%%
%%%%%%%%%%%%%%% Coherent response and Incoherent cavity Noise %%%%%%%%%%%%%%%%%%%%%
%%%%%%%%%%%%%%%%%%%%%%%%%%%%%%%%%%%%%%%%%%%%%%%%%%%%%%%%%%%%%%%%%%%%%%%%%%%%%%%%%%%%%%%%%%%%

\section{Coherent response and Incoherent cavity Noise}

As the thermally driven resonance shift is given by:
\begin{equation}
    \frac{\delta n}{n} = -\frac{\int \delta T(\vec{r}) \varepsilon_0 \sqrt{\varepsilon_r(\vec{r})}\beta |u(\vec{r})|^2 \ \mathrm{d}^3 r}{\int  \varepsilon_0 \varepsilon_r |\vec{u}(\vec{r})|^2\ \mathrm{d}^3 r},
\end{equation}
From the fluctuation-dissipation-theorem introduced in the previous section, the themo-refractive noise at given offset frequency $f = 2\pi \omega $ is given by the dissipated energy $W_{diss}$ when the system is periodically injected by a entropy flow:
\begin{equation}
    q(\vec{r},t)= -\frac{ \varepsilon_0 \sqrt{\varepsilon_r(\vec{r})}\beta |u(\vec{r})|^2 }{\int  \varepsilon_0 \varepsilon_r |\vec{u}(\vec{r})|^2\ \mathrm{d}^3 r}\ \cos(\omega t)
\end{equation}

The frequency dependency of the thermorefractive noise $S_{\nu,TRN}(\omega)$, is thus given by the thermal diffusion process with a spatial heat-source distribution following $q(\vec{r},t)$. 
In coherent response measurement, the pump laser is periodically intensity modulated by the intensity modulator, and the photothermal absorption of the pump laser results in a spatial distributed heat-source following:
\begin{equation}
    q_{\mathrm{coherently\ driven}} \propto \alpha_{\mathrm{ab.}} \frac{ \varepsilon_0 \sqrt{\varepsilon_r(\vec{r})} |u(\vec{r})|^2 }{\int  \varepsilon_0 \varepsilon_r |\vec{u}(\vec{r})|^2\ \mathrm{d}^3 r}\ \cos(\omega t)
\end{equation}
where $\alpha_{\mathrm{ab.}}$ is the photothermal absorption coefficient, which depends on the material and fabrication process. 
It can be seen that in the coherent driven response measurement, the pump induced photothermal heat source $q_{\mathrm{coherently driven}}$ has similar spatial distribution as the imaginary heat source we introduced in the calculation of thermorefractive noise based on fluctuation-dissipation theorem:
\begin{equation}
    q(\vec{r,t}) \sim q_{\mathrm{coherently\ driven}}
\end{equation}

and the frequency dependency of the thermorefractive noise and coherent response are both determined by the thermal dissipation process, thus yields to the same frequency scaling, and also acts as a tool to distinguish that the measured cavity noise is governed by the thermal diffusion process or not. In our measurement, the cavity noise in the non-electro-optic material, silicon nitride, is thermorefractive noise that governed by the thermal diffusion process, thus the coherently driven response measurement exhibits the same frequency scaling as the incoherent noise measurement. However, the electro-optic material we studied, including the Lithium Niobate and Lithium Tantalte, the charge-carrier noise is determined by the charge-diffusion process, rather than the thermal diffusion process, and thus the coherent driven response measurement deviates from the incoherent cavity noise measurement.

\begin{figure*}[htb]
	{
		\centering
		\includegraphics[width=0.8\linewidth]{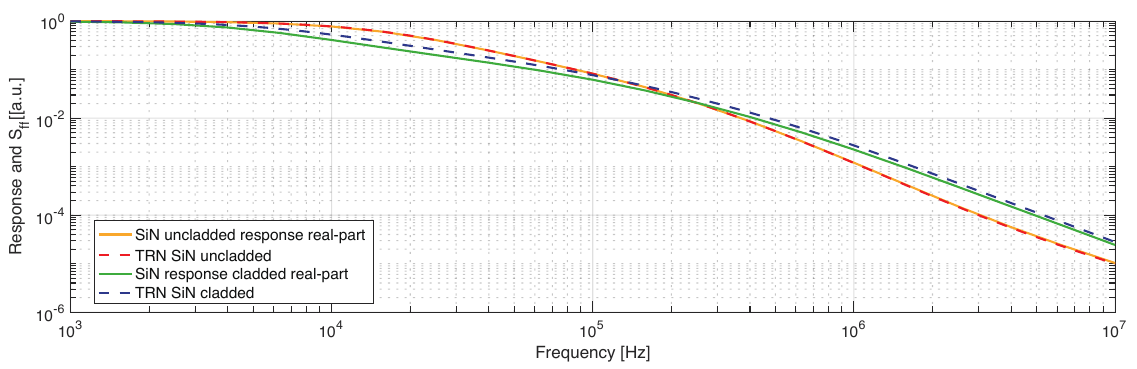}
		\caption{\textbf{Simulation result for cavity noise and response.}
			The simulation results show the thermo-response and thermorefractive noise in cladded and uncladded silicon nitride microresonators. The real-part of the cavity-driven thermo-response exhibits the same frequency scaling as the thermorefractive noise, in accordance with the fluctuation-dissipation theorem.
		} 
		\label{TRN_and_response}       
	}
\end{figure*}

\subsection{Experiment Setup for Coherently-Drived Nonlinear Response Measurement}
\begin{figure*}[htb]
    {
        \centering
        \includegraphics[width=1.0\linewidth]{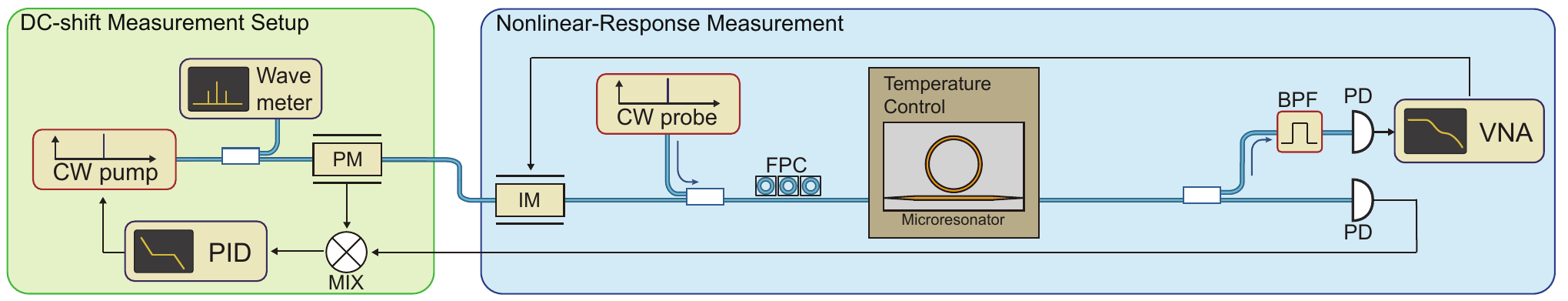}
        \caption{\textbf{Experiment setup for DC-shift measurement and coherent nonlinear response measurement.}
            CW pump/probe: continuous wave external cavity diode laser;
            PM: phase modulator;
            IM: intensity modulator;
            BPF: band-pass filter;
            FPC: fiber polarization controller;
            MIX: microwave mixer;
            PID: proportional–integral–derivative controller;
            VNA: vector-network analyzer.
            DC-shift measurement: A pump laser is PDH-locked to the cavity under test and monitored by a wavelength meter to track the temperature-induced resonance shift. 
            The photonic chip temperature is controlled by a thermoelectric cooler. 
            Nonlinear response measurement setup: The pump laser is intensity modulated by an intensity modulator to drive a nonlinear process in the cavity.
            A probe laser is then placed on the side of the resonance to probe the non-linear shift induced by the modulated pump.  The phase and modulation depth of the resulting response is recorded using a VNA.
        } 
        \label{setup_response}       
    }
\end{figure*}

The experiment setup for coherent nonlinear response measurement to probe thermo-optic response is shown in \fref{setup_response}.
\subsection{Additional Data of Coherent Response Measurement}
\begin{figure*}[htb]
	{
		\centering
		\includegraphics[width=1.0\linewidth]{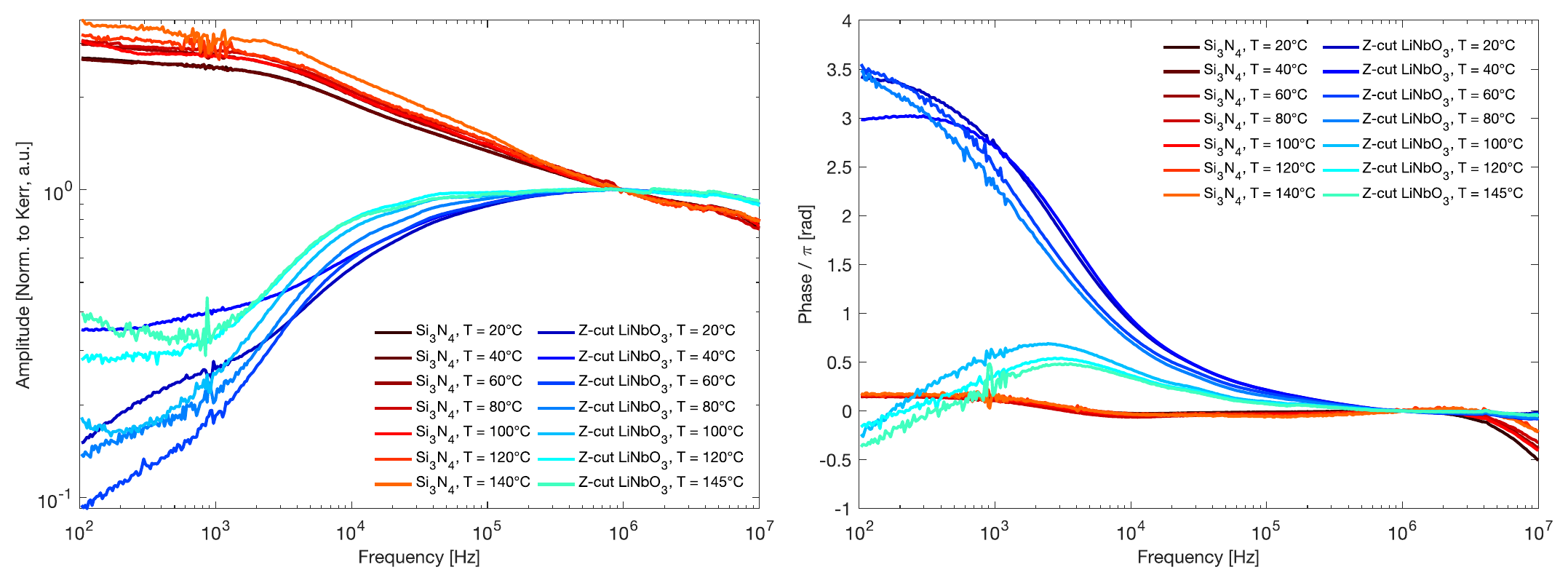}
		\caption{\textbf{Coherent response measured in different temperature.}
            The coherently driven non-linear response (normalized to the Kerr nonlinearity) in different temperature from 20$\mathrm{^\circ C}$ to 140$\mathrm{^\circ C}$. The measured results are normalized to the Kerr-nonlinearity at 1MHz offset frequnecy.
		} 
		\label{different_T_response}       
	}
\end{figure*}

\fref{different_T_response} plots the coherently driven non-linear response measured in \SiN and Z-cut \LN microresonators at various temperatures.
As the temperature increases, the photothermal nonlinearity $|\gamma|$ slightly increases in \SiN and decreases in Z-cut \LN. Due to the PyroEO-induced negative themo-optic coefficient which over-compensates the intrinsic material thermo-optic coefficient, the Z-cut \LN exhibits an overall negative thermo-optic coefficient which shifts the cavity resonance in the opposite direction when compared with the Kerr nonlinearity.

At room temperature, the photothermal nonlinearity is relatively large ($|\gamma| > 1$), causing an over-compensation of the Kerr nonlinearity and leading to a $\pi$ phase shift. As the temperature increases, the photothermal nonlinearity weakens to the point where $|\gamma| < 1$, and it can no longer surpass the Kerr nonlinearity. As a result, no $\pi$ phase shifts are observed

%%%%%%%%%%%%%%%%%%%%%%%%%%%%%%%%%%%%%%%%%%%%%%%%%%%%%%%%%%%%%%%%%%%%%%%%%%%%%%%%%%%%%%%%%%%%
%%%%%%%%%%%%%%%%%%%%%% Long-time drift in DC-drift measurement %%%%%%%%%%%%%%%%%%%%%%%%%%%%%
%%%%%%%%%%%%%%%%%%%%%%%%%%%%%%%%%%%%%%%%%%%%%%%%%%%%%%%%%%%%%%%%%%%%%%%%%%%%%%%%%%%%%%%%%%%%
\section{Long-time drift in DC-drift measurement}
\subsection{Hysteresis in of the PyroEO effect}
\noindent In the main text, we presented experimental results demonstrating a hysteresis and slow relaxation in the temperature-induced resonance frequency shift of Z-cut \LN-based optical microresonator. 
We observed in Figure 2(b) of the main manuscript that when the temperature was changed at a rate of approximately 0.5K/min, the data points exhibited a characteristic shape that could be effectively fitted using a quadratic function.
In this section, we will provide a detailed explanation for this observed effect, attributing it to the presence of a slow-decay of the pyroelectric charges.
The electric field $E$ generated by the pyroelectric effect is proportional to the temperature $T$. 
However, it is important to note that the electric field experiences a decay rate $\gamma$ due to the presence of environmental charge neutralization and the leakage current through the \LN crystal (z-cut) and along its surface (x-cut)
\begin{equation}
	\frac{\mathrm{d}E}{\mathrm{d}t} = -\gamma E + \alpha \frac{\mathrm{d}T}{\mathrm{d} t}.
\end{equation}
Thus we have:
\begin{equation}
	\begin{aligned}
		E(t) = E(0)e^{-\gamma t} + \alpha\int_0^t\ \frac{\mathrm{d}T(\tau)}{\mathrm{d} \tau}e^{-\gamma(t-\tau)}\ \mathrm{d}\tau.
	\end{aligned}
\end{equation}
The typical time scale for DC-shift measurement is generally much smaller than the characteristic time for decay $1/\gamma$, which is confirmed by our ring-down experiment presented in Figure 2(d) of the main manuscript. 
Therefore, we can express this relationship as:
\begin{equation}
	\begin{aligned}
		E(t) &\approx E(0)e^{-\gamma t} + \alpha\int_0^t\ \frac{\mathrm{d}T(\tau)}{\mathrm{d} \tau} \left(1-\gamma\left(t-\tau\right)\right)\ \mathrm{d}\tau  \\     
		&= E(0)e^{-\gamma t} + \alpha (T(t) - T(0)) - \gamma\alpha\int_0^t\ \frac{\mathrm{d}T(\tau)}{\mathrm{d} \tau}(t-\tau)\ \mathrm{d}\tau \\ 
		&= E(0)e^{-\gamma t} + \alpha (T(t) - T(0)) + \gamma\alpha t T(0) + \gamma \alpha \int_0^t\ T(\tau)\ \mathrm{d}\tau.
	\end{aligned}
\end{equation}
Hence we find that a linear temperature ramp with a constant rate of change, expressed as $\frac{\mathrm{d}T}{\mathrm{d}\tau}=Const.$, along with an initial condition of $E(0)=0$, leads to a quadratic time dependence of the resonance frequency shift commensurable with the experimental observation.

\subsection{Extended data for long-time DC-shift measurement}
\begin{figure*}[htb]
	{
		\centering
		\includegraphics[width=0.8\linewidth]{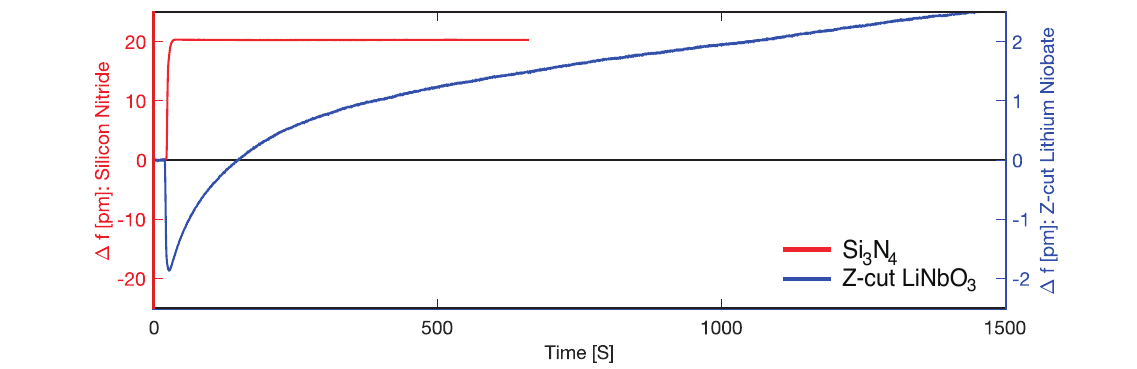}
		\caption{\textbf{Time-dependency of the temperature induced frequency shift.}
			The temperature is ramped down from 22~$^\circ C$ to 23~$^\circ C$ in 10 seconds and the resonance shift for TE-mode in both silicon nitride and Z-cut Lithium Niobate is continously monitored over a duration of 1500~s.
		} 
		\label{long_ring}       
	}
\end{figure*}

\noindent The longer duration of the ring-down experiment, as depicted in SI \fref{long_ring}, demonstrates the progressive shift of the resonance of the z-Cut \LN-based optical microresonator due to the photo-refractive effect, as the resonator is continuously illuminated over a period of 1500~s after the temperature is initially ramped from 22$^\circ$C to 23$^\circ$C in 10 seconds. 
The laser power during the experiment coupled into the optical waveguide was fixed as $3\mu W$.
We note that the z-cut \LN-based optical microresonator initially exhibits an effective negative thermo-optic coefficient as a result of the PyroEO effect, causing a red shift of the resonance as the temperature decreases. 
However, after the temperature stabilizes, the resonance gradually shifts towards the blue side due to the gradual dissipation of pyroelectric charges and the presence of the photorefractive effect. 

%%%%%%%%%%%%%%%%%%%%%%%%%%%%%%%%%%%%%%%%%%%%%%%%%%%%%%%%%%%%%%%%%%%%%%%%%%%%%%%%%%%%%%%%%%%%
%%%%%%%%%%%%%%% Experiment details in Intrinsic Noise Measurement %%%%%%%%%%%%%%%%%%%%%%%%%%
%%%%%%%%%%%%%%%%%%%%%%%%%%%%%%%%%%%%%%%%%%%%%%%%%%%%%%%%%%%%%%%%%%%%%%%%%%%%%%%%%%%%%%%%%%%%
\subsection{Experiment details in Intrinsic Noise Measurement}
\begin{figure*}[htb]
    {
        \centering
        \includegraphics[width=1\linewidth]{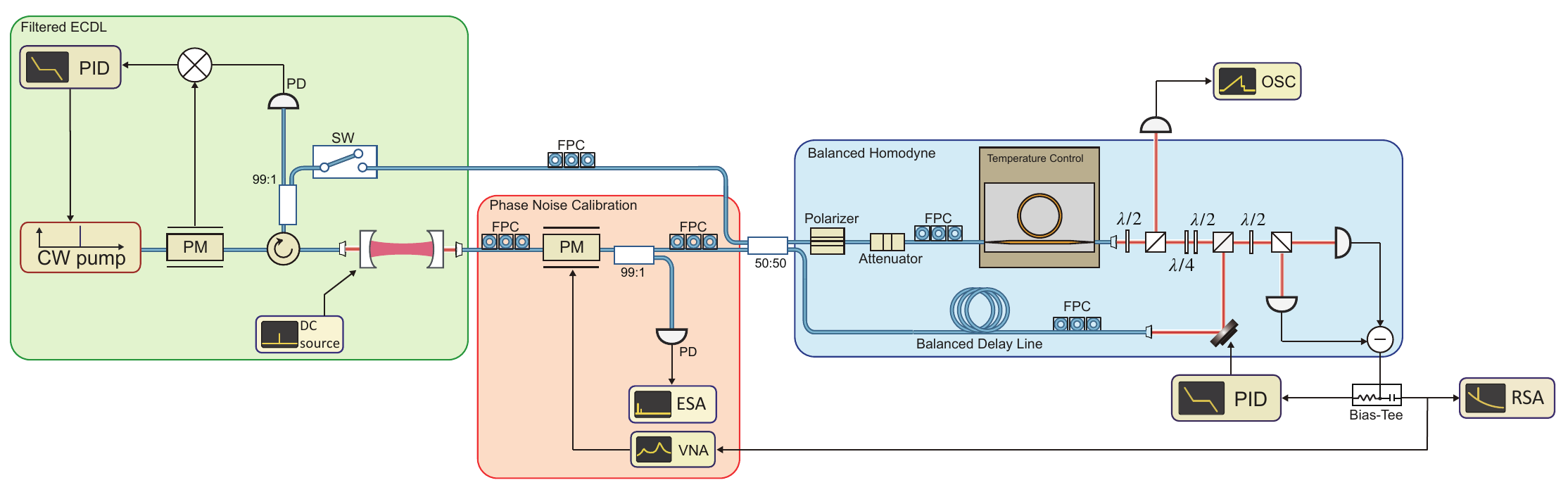}
        \caption{\textbf{Complete experiment setup for intrinsic cavity noise measurement.}
        The overall setup can be divided into three main components:
        1. Filtered External Cavity Diode Laser (ECDL): An external cavity diode laser is locked to a narrow linewidth Fabry-Perot cavity using the Pound-Drever-Hall (PDH) technique. This configuration effectively filters out high-frequency noise with offsets larger than the linewidth (70~kHz) of the Fabry-Perot cavity.
        2. Balanced Homodyne Detection: A balanced Mach-Zehnder interferometer (MZI) is locked to the phase quadrature by a piezo-mirror. 
        The intrinsic frequency noise of the cavity is converted into laser phase noise as it passes through the cavity. 
        This phase noise is then measured using the balanced homodyne detection technique.
        3.Phase Noise Calibration. Phase Noise Calibration: A phase modulator is employed to generate a calibration peak for measuring the phase noise. Polarization monitoring is implemented to minimize the effects of remaining amplitude modulation (RAM), and a vector network analyzer is utilized to monitor RAM levels.
        } 
\label{detailed_experiment_TRN}       
    }
\end{figure*}

The experimental setup for measuring intrinsic cavity noise is depicted in \fref{detailed_experiment_TRN}. The setup has been adapted and modified from a previous work by Huang et al \cite{huangThermorefractiveNoiseSiliconnitride2019}.

\noindent The measurement principle employed in this experiment is phase detection based on a balanced homodyne interferometer.
A laser is precisely tuned to the resonance frequency of the cavity being tested, resulting in a phase shift $\phi(\Delta)$ that depends on the relative detuning $\Delta = \omega - \omega_0$ between the laser frequency $\omega$ and the resonance frequency of the cavity $\omega_0$. 
This process enables the conversion of the intrinsic frequency noise of the cavity into phase noise on the laser beam.
The phase noise is further converted into intensity noise through interference. 
This is achieved by employing a piezo-mirror to lock the homodyne to the phase quadrature. 
By tuning the optical path lengths of the local oscillator and signal branches to be equal when the laser is off-resonance, a balanced homodyne configuration is created. 
This configuration effectively suppresses the intensity noise from the measurement laser, ensuring accurate measurement of the converted phase noise as intensity fluctuations.
To achieve a low-frequency noise laser, we employ a PDH lock between an external cavity diode laser (ECDL) and a high-finesse Fabry-Perot cavity (linewidth < 70~kHz). The transmitted light from the Fabry-Perot cavity is used to drive the homodyne detection system.
To maximize the transduction of cavity noise into phase noise of the transmitted laser, it is crucial to tune the laser to resonance by setting $\Delta = 0$. This is accomplished by incorporating a piezo element onto the Fabry-Perot (FP) cavity, which is used to lock the low-noise cavity to the optical resonance with a low-pass filter cut-off of 100~Hz allowing for accurate measurement of noise at offset frequencies larger than 1~kHz.
Prior to locking, the reflected light from the FP cavity is employed to align the resonance of the FP cavity with that of the microresonator being tested. This alignment ensures that the laser is precisely tuned to the resonance of the microresonator, enabling efficient transduction of cavity noise into phase noise.

\begin{figure*}[htb]
    {
        \centering
        \includegraphics[width=0.8\linewidth]{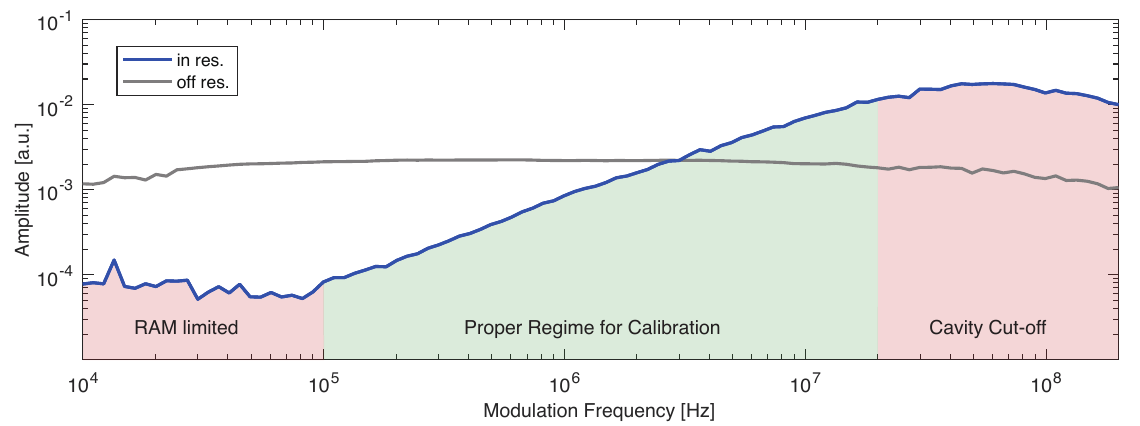}
        \caption{\textbf{Response function of the phase calibration peak.}
        The typical measurement result obtained from a vector network analyzer (VNA) for the phase calibration peak, with the laser operating both in resonance and off resonance, reveals distinct response patterns. These patterns can be categorized into three regions: The response can be seperated into three regions: 
        1. Residual amplitude modulation limited: When the calibration frequency is too low, the dominant contribution to the phase modulation response is the residual amplitude modulation (RAM), making it unsuitable for use as a reliable calibration peak. 
        2. Proper Regime for Calibration: As the modulation frequency increases, the phase modulation becomes dominant over the residual amplitude modulation (RAM) due to the proportional relationship between the frequency modulation depth and modulation frequency for a given phase modulation depth. Consequently, this results in a proper calibration and accurate measurements.
        3. Cavity cut-off regime: when the modulation frequency is high enough to approach the linewidth of the cavity under-test, the cavity filters the generated sideband leading to inaccurate calibration.
        Therefore, it is crucial to ensure that the calibration peak is positioned within the appropriate range during the experiment.
        } 
\label{TRN_RAM_VNA}       
    }
\end{figure*}

To calibrate the frequency noise measurement, a phase modulator is employed before the homodyne detection system. 
This phase modulator introduces a phase modulation peak with known frequency and depth. 
The noise spectra obtained from the real-time spectrum analyzer (RSA) is then calibrated using this known calibration peak. 
This calibration allows us to extract the extraction of the cavity noise power spectral density denoted as $S_{\nu}(f)$.
We characterize the phase noise calibration by inserting a Vector Network Analyzer into the system and we investigate the frequency response of the phase modulator and balanced homodyne system (cf. \Fref{TRN_RAM_VNA}).
The presence of residual amplitude modulation (RAM) at low offset frequencies in the phase modulator is an undesired effect which can impact calibration. 
We note that the low frequency deviation is not due to the 9~kHz bias-tee that we use to separate the PID lock error signal from the homodyne phase signal. 
At high offset frequencies the optical microresonator filters the phase modulation sideband (cavity Cut-off), again leading to unreliable calibration. 
During the experiment, the polarization at the phase modulator input is controlled precisely to minimize RAM and the calibration peak is positioned at a frequency that is sufficiently high to avoid RAM while still being lower than the cavity cut-off frequency.

\subsection{Additional Data for Intrinsic Cavity Noise Measurement}
\begin{figure*}[htb]
    {
        \centering
        \includegraphics[width=1\linewidth]{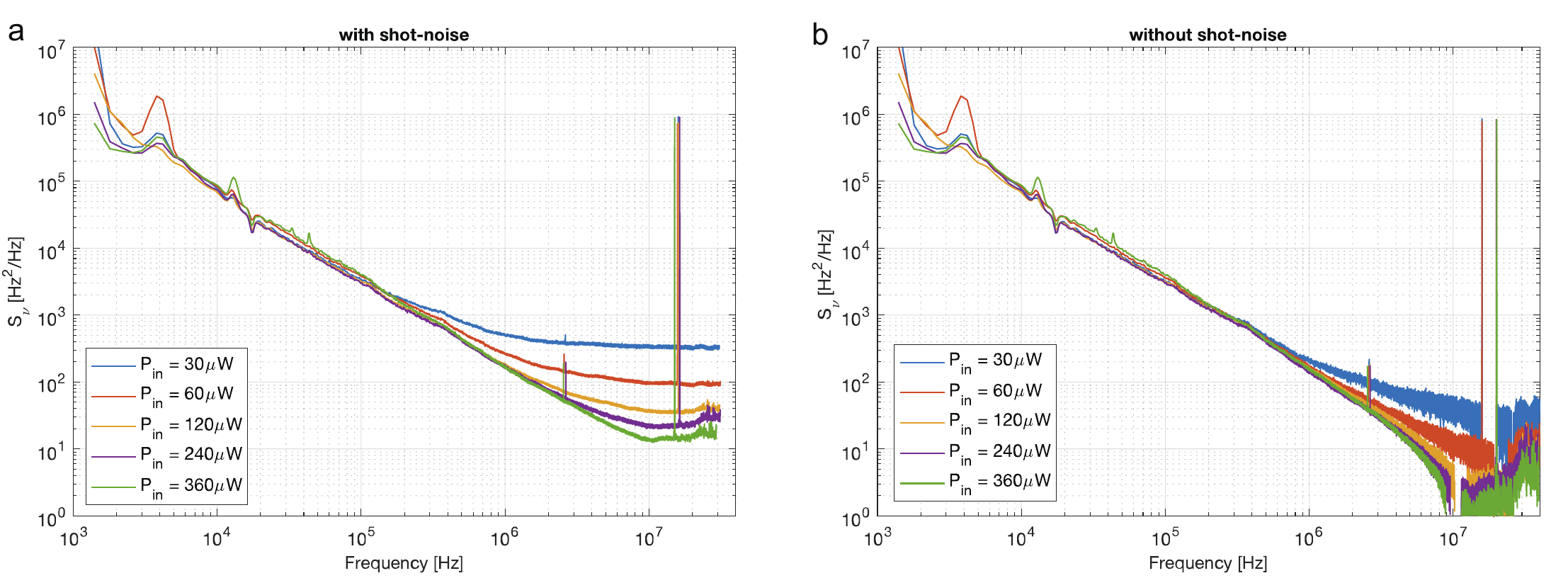}
        \caption{\textbf{Intrinsic cavity noise measurement on 200GHz Z-cut \LN microresonator with various input power.}
        The intrinsic cavity frequency noise of a 200GHz Z-cut \LN microresonator is measured at various input powers $P_{in}$ on the input-fiber. The fiber-chip coupling is approximately -5dB. In (a), the data is presented without subtracting the shot-noise level, while in (b), the shot-noise removed data is displayed.
        } 
\label{various_power}       
    }
\end{figure*}

\fref{various_power} plots the intrinsic cavity noise measurement result in 200GHz Z-cut \LN microresonator with various input power. The input power is measured on the input coupling fiber, and due to the coupling loss in the fiber-to-chip interface, the power on the bus waveguide will be approximate 5dB lower. 

The local-oscillator power is kept unchanged during the measurement and is significantly larger than the signal power, the result is limited by the optic shot-noise. With the increase signal power, the signal-to-noise ratio is increased. 
To correctly address the intrinsic cavity noise in high-offset frequency, the shot-noise level must be subtracted from the measured result.

In the power range we measured from 30$\mathrm{\mu W}$ to 360$\mathrm{\mu W}$, as shown in \Fref{various_power}, no significant different in the measured cavity noise can be seen.

\begin{figure*}[htb]
    {
        \centering
        \includegraphics[width=0.7\linewidth]{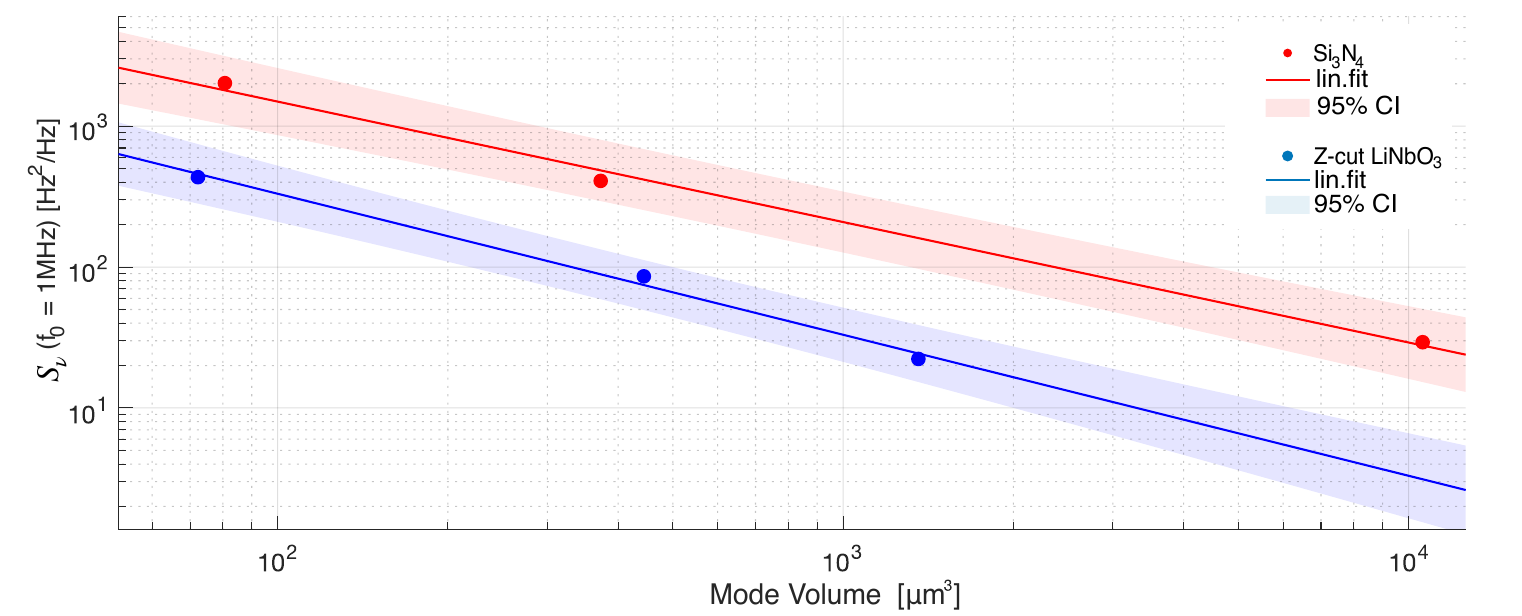}
        \caption{\textbf{Intrinsic cavity noise measurement scales with mode-volume.}
        Measured single-sided intrinsic single-sided cavity frequnecy noise at offset frequency $f_0 = 1\mathrm{MHz}$ in Z-cut \LN and \SiN microresonators with various free-spectra range (scales to mode-volume).
        } 
\label{various_volume}       
    }
\end{figure*}

\Fref{various_volume} plots the measured intrinsic single-sided cavity frequency noise at offset frequency $f_0=1\mathrm{MHz}$ in Z-cut \LN and \SiN microresonators, with various free-spectra range and scales to mode-volume $V_{\mathrm{eff}}$. The mode volumes are calculated using finite-element methods based on the geometry of the waveguide crosssection. \Fref{various_volume} shows that the cavity noise is consistently lower in Z-cut\LN than in \SiN, and is inversely proportional to the mode-volume within the confidence interval: $S_{\nu}\propto V_{\mathrm{eff}^{-1}}$, aligned with our theory.

\section{Self-Injection Locking Measurement}

\noindent Self-injection locking is a commonly employed technique for constructing narrow linewidth integrated lasers. 
A comprehensive review detailing recent advancements in this field can be found in Kondratiev et al. (2023) \cite{kondratievRecentAdvancesLaser2023}. 
The fundamental working principle of self-injection locking is straightforward. 
It involves utilizing an external microresonator cavity with a high quality factor, which generates a sharp reflection peak at resonance, surpassing the sharpness of the resonance peak in the laser cavity itself. As a result, the frequency stabilization is significantly enhanced. 
Consequently, the laser's linewidth and frequency noise are proportionally reduced, with the reduction being determined by the square of the stabilization coefficient  \cite{laurentFrequencyNoiseAnalysis1989,spanoTheoryNoiseSemiconductor1984}. 

In our case, a distributed feed-back laser (DFB) is self-injection locked to the on-chip microresonator, and the reflection is relatively weak in order to avoid instability, the frequency noise reduction factor of the locked laser can be written as \cite{kondratievSelfinjectionLockingLaser2017}:
\begin{equation}
    \frac{\delta f}{\delta f_{free}} \approx \frac{Q_d^2}{Q_{m}^2}\frac{1}{16\Gamma_m^2(1+\alpha_g^2)} \propto \frac{1}{Q_m^2},
\end{equation}
where $\delta f_{free}$ is the frequency fluctuation of the DFB laser due to the drive current noise and temperature fluctuation; the $\delta f$ is the frequency deviation of the whole hybrid laser after self-injection locking; $Q_d\sim 10^3-10^4$ is the quality factor of the DFB laser diode, $Q_{m}\sim 10^6-10^7$ is the quality factor of the microresonator; $\Gamma_m$ describes the amplitude reflection coefficient from the microresonator at resonance and $\alpha_g$ is the phase-amplitude coupling factor, which is of the order of unity.  

\begin{figure*}[htb]
    {
        \centering
        \includegraphics[width=1\linewidth]{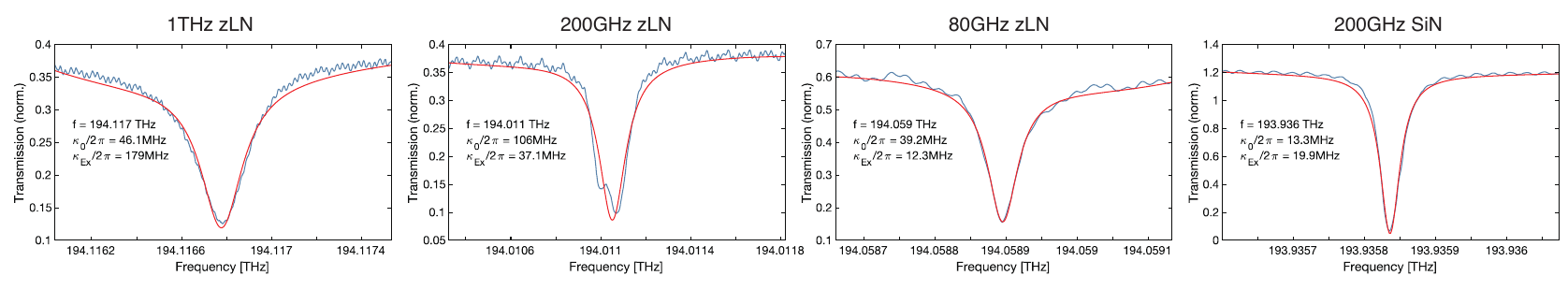}
        \caption{\textbf{Resonances for self-injection locking experiment.}
        The cavity resonances that used for self-injection-locking experiment. $\kappa_0$ and $\kappa_{ex}$ are intrinsic cavity dissipation and external coupling strength, respectively, determined through phase shift measurements. The presence of a $\pi$ phase-shift near the resonance helps identify if the system is overcoupled or undercoupled.
        } 
\label{SIL_res}       
    }
\end{figure*}

\fref{SIL_res} plots the resonances at which our self-injection locking experiment is perform on.
From the equation mentioned above, it is evident that the frequency noise of the hybrid laser decreases with an increase in the quality factor of the microresonator.

However, it should be noted that the presence of intrinsic cavity noise in the microresonator, arising from factors such as thermorefractive noise or charge-carrier noise in this case, imposes a limitation on the minimum frequency linewidth as it is not reduced by self-injection locking but directly imprints on the frequency fluctuation of the hybrid laser. 
Hence, despite the decrease in frequency noise with an increase in the microresonator's quality factor, the intrinsic cavity noise sets a lower bound on the achievable linewidth.
\begin{equation}
    S_\mathrm{\nu,\ hybrid\ laser} = \min \left[ \frac{Q_d^2}{Q_{m}^2}\frac{1}{16\Gamma_m^2(1+\alpha_g^2)}  S_\mathrm{\nu,\ free},\ S_\mathrm{\nu,\ cavity\ noise} \right],
\end{equation}
where $S_{ff, cavity\ noise}$ refers to the intrinsic cavity noise. 
In the ferroelectric material-based microresonators we investigated, our findings in the main text indicate that the presence of pyroelectricity and charge-carrier noise leads to a cavity noise, denoted as $S_\mathrm{\nu, cavity\ noise}$, that is lower than the predicted value based on conventional thermorefractive noise. 
Moreover, the measured cavity noise exhibits a Flicker-like behavior.
The high-quality factor of our microresonator enables the hybrid laser to attain the modified cavity noise limit in terms of frequency noise.

\subsection{Frequency noise measurement by beat-note}
\noindent The laser frequency noise of our hybrid laser is characterized using heterodyne spectroscopy with a reference laser, specifically a cavity-filtered external cavity diode laser. 
It is important to ensure that the frequency noise of the reference laser is lower than that of the hybrid laser being characterized, such that:
\begin{equation}
    \begin{aligned}
        S_\mathrm{\nu,\ mea} &= S_\mathrm{\nu,\ hybrid\ laser} + S_\mathrm{\nu,\ ref.\ laser}\\
         &\approx S_\mathrm{\nu,\ hybrid\ laser}\ (\mathrm{if}\ S_\mathrm{\nu,\ hybrid\ laser} \gg S_\mathrm{\nu,\ ref.\ laser})
    \end{aligned}
\end{equation}

\begin{figure*}[htb]
    {
        \centering
        \includegraphics[width=1\linewidth]{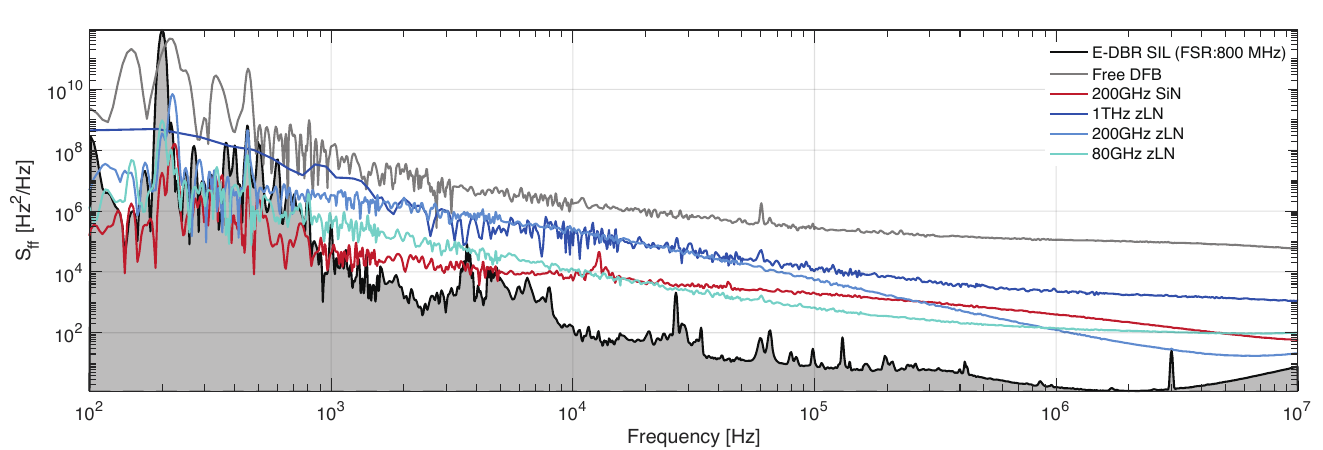}
        \caption{\textbf{Frequency noise spectra for self-injection locking experiment.}
       The power spectral density (PSD) of the heterodyne beat-note between different self-injection locked lasers and the reference laser is presented. The PSD obtained for the E-DBR SIL, as adopted from \cite{siddharthHertzlinewidthFrequencyagilePhotonic2023}, serves as an upper-limit estimate for the frequency noise of the reference laser.
        } 
\label{toptica_noise}       
    }
\end{figure*}
\fref{toptica_noise} shows the beat-note frequency noise $S_{\nu,\ mea}$ between reference laser and various hyrbid lasers. It is obvious that the beat-note noise for all four hybrid laser experiment in our paper is all above the beat-note noise for EDBR-based hybrid laser \cite{siddharthHertzlinewidthFrequencyagilePhotonic2023}, which acts as an upper limit for the frequnecy noise of the reference laser.
That means $S_{ff,\ hybrid\ laser} \gg S_{ff,\ ref.\ laser}$ is satisfied in our experiments, thus the beat-note based frequency noise measurement is valid.

\bibliography{refs.bib}